\documentclass{mn2e}

\usepackage{subfigure} \usepackage{graphicx} \usepackage{setspace}
\usepackage[dvips]{epsfig}

\def\be{\begin{equation}} 
\def\ee{\end{eqnarray}}
\def\ben{\begin{eqnarray}} 
\def\een{\end{equation}}
 
\def\omm{\Omega_m} 
\def\oml{\Omega_\Lambda}

\def\s8{\sigma_8} 

\def\hMsol{h^{-1}\,\hbox{M}_{\odot}} 
\def\dlnsiginv{d \mbox{ln}\sigma^{-1}} 
\def\lnsiginv{\mbox{ln}\sigma^{-1}} 
\def\dlnM{d \mbox{ln} M}
\def\hvect#1{{\hat{\mathbfit{#1}}}}
\def\kmsMpc{\,\hbox{km}\,\hbox{s}^{-1}\,\hbox{Mpc}^{-1}}
\def\hMpc{\,h^{-1}\,\hbox{Mpc}}

\begin{document} 
\title[Sunyaev-Zel'dovich Cluster Survey Simulations for {\it Planck}]     
      {Sunyaev-Zel'dovich Cluster Survey Simulations for {\it Planck}}
\author[J.~Geisb\"{u}sch, R.~Kneissl, M.~Hobson]  
       {J\"{o}rn~Geisb\"{u}sch\thanks{email: joern@mrao.cam.ac.uk},
	R\"{u}diger~Kneissl and Michael~Hobson\\ Astrophysics Group, 
	Cavendish Laboratory, Madingley Road, Cambridge, CB3 0HE, UK}
\maketitle
\begin{abstract}
We examine the ability of the future Planck mission to provide a
catalogue of galaxy clusters observed via their Sunyaev-Zel'dovich
distortion in the cosmic microwave background.  For this purpose we
produce full-sky Sunyaev-Zel'dovich maps based on N-body simulations
and scaling relations between cluster properties for several
cosmological models. We extrapolate the N-body simulations  by a mass
function to high redshifts in order to obtain a realistic SZ
background. The simulated Planck observations include, besides the
thermal and kinematic Sunyaev-Zel'dovich effects, contributions from
the primordial CMB, extragalactic point sources as well as Galactic
dust, free-free and synchrotron emission. A harmonic-space
maximum-entropy method is used to separate the SZ signal from
contaminating components in combination with a cluster detection
algorithm based on thresholding and flux integration to identify
clusters and to obtain their fluxes. We estimate a survey sensitivity
limit (depending on the quality of the recovered cluster flux) and
provide cluster survey completeness and purity estimates. We find that
given our modelling and detection algorithm Planck will reliably
detect at least several thousands of clusters over the full sky. The
exact number depends on the particular cosmological model (up to 10000
cluster detections in a concordance $\Lambda$CDM model with $\sigma_8
= 0.9$). We show that
the Galaxy does not significantly affect the cluster
detection. Furthermore, the dependence of the thermal SZ power
spectrum on the matter variance on scales of 8 $h^{-1}$Mpc and the
quality of its reconstruction by the employed method are investigated.
Our simulations suggest that the Planck cluster sample will not only
be useful as a basis for follow-up observations, but also will have
the ability to provide constraints on cosmological parameters.
\end{abstract}


\section{Introduction}
\label{sec:intro}

In addition to measuring primordial CMB fluctuations, the Planck Surveyor satellite, which is scheduled for launch in 2007, will
have the ability to construct a cluster catalogue via the
Sunyaev-Zel'dovich (SZ) effect (Sunyaev \& Zel'dovich 1972; Sunyaev \&
Zel'dovich 1980; recent reviews: Rephaeli 1995; Birkinshaw 1999; Carlstrom, Holder, \& Reese 2002). Since Planck will provide
detailed full-sky maps at nine different frequencies, ranging from 30
to 850 GHz, it is a particularly suitable instrument to detect the
thermal SZ effect of galaxy clusters owing to its distinctive frequency
dependence. Moreover, since the SZ effect is redshift independent and
clusters -- as the most massive collapsed objects in the universe --
are tracers of the underlying linear matter density perturbations, their
observation provides a unique tool for studying large scale structure
and constraining cosmological parameters.

In this work we investigate the yield, completeness and purity of such
a Planck SZ survey for a
range of cosmological models and estimate the cosmological
significance of such the resulting cluster catalogue. Furthermore, based on
our simulations and cluster selection we estimate a Planck cluster
survey flux limit and provide tabulated cluster number counts for each model. In
contrast to previous work (Kay, Liddle, \& Thomas 2001; Diego et
al.\,2002; White 2003), which studied the cluster sample
either on sky patches or neglected contaminants, we perform full-sky
simulations including all known significant extragalactic and Galactic
contaminating components. Our modelling of the SZ effect as well as
contaminants aims to be as realistic as current observational
constraints allow. To recover the SZ signal of single clusters we make
use of the harmonic-space maximum entropy method introduced by
Stolyarov et al.\,(2002) in combination with a cluster finding algorithm
based on peak detection, thresholding and flux integration. As Stolyarov et
al.\,(2002) show the HSMEM is a novel method to separate various CMB
components. The cluster finding algorithm is shown in this paper to be
able to obtain reliable
flux as well as radius estimates for a cluster sample above the
estimated survey flux limit. Note that besides the
assumed cosmological model the cluster recovery algorithm has major
implications on the number of clusters obtained by the Planck survey
and thus its impact on constraining cosmological parameters. It is
therefore important to explore,
improve and combine techniques in order to find the optimal method and
maximise the survey yield. Since our modelling exhibits a high degree
of realism, the given estimate of the cluster number which can be
obtained by the Planck survey using our algorithm is reliable. Even
though further improvement is
possible, the presented combined cluster detection method -- as we
show below -- provides already the quality to draw interesting cosmological
conclusions from the obtained cluster sample.

The outline of the paper is as follows. In section \ref{sec:sztheo} a short
introduction to the SZ effect is given. The modelling of full-sky SZ
maps of different cosmological models and of the contaminants is
described in section \ref{sec:szsim}. Moreover,
section \ref{sec:szsim} contains a brief description of how Planck observation
simulations are obtained from the modelled components. In section
\ref{sec:recon}, the applied reconstruction method used to recover the
fluxes of single galaxy clusters is explained. Section
\ref{sec:resanddis} contains a presentation and discussion of the
results. Finally, a summary and conclusions are given in section
\ref{sec:concl}.

\section{The Sunyaev-Zel'dovich effect}
\label{sec:sztheo}

The CMB anisotropy caused by the SZ effect can be separated into two
contributions which are distinguished by the origin of energy of the
scattering electrons that is responsible for the shift of photon
frequency. The total CMB distortion due to the SZ effect is given by
\begin{equation}
\frac{\Delta I_{\nu}}{I_0} = g(x) y - h(x)\beta \tau \,,
\label{equ:sz}
\end{equation}
where $x=h\nu/k_{{\rm B}}T_0$ with $T_{0}=2.725 \, {\rm K}$
(Mather et al.~1999) and $I_0 =2k_{{\rm B}}^3T_0^3/h^2c^2$.
The first term in equation (\ref{equ:sz}) is the so called thermal SZ
effect due to the thermal motion of electrons of the intra-cluster
gas. The thermal SZ effect has a spectral shape given by
\begin{equation}
g(x) = {x^4 e^{x} \over (e^{x}-1)^2} \, \left[ x {e^{x}+1 \over
e^{x}-1} - 4 \right],
\label{equ:tszspec}
\end{equation}
and a frequency independent magnitude, the Comptonization parameter,
\begin{equation}
y = {k_{{\rm B}} \sigma_{\rm T} \over m_e c^2} \, \int \, n_e T_e \, dl \,.
\label{equ:tszy}
\end{equation}
In hot clusters ($T_e > 5$keV) the relativistic electrons present
slightly modify the spectral shape of the thermal SZ effect (Challinor
\& Lasenby 1998). This resulting relativistic correction has not been
taken into account in this work, since its effect on
the results presented is negligible. The detectability of the effect
from thermal (e.g. Pointecouteau et al.\,1998) and non-thermal (e.g. Ensslin \&
Hansen 2004) relativistic electrons has been
estimated. The spectral shape of the second contribution in
equation (\ref{equ:sz}), the kinematic SZ effect, is given by
\begin{equation}
h(x) = {x^4 e^{x} \over (e^{x}-1)^2},
\label{equ:kszspec}
\end{equation}
and its magnitude, $\beta = v_{\rm pec}/c$, depends on the uniform
peculiar line-of-sight bulk motion of the cluster's electron plasma,
$v_{\rm pec}$.
\begin{equation}
\tau = \sigma_T \, \int \, n_e \, dl \,,
\label{equ:thoptdepth}
\end{equation}
is the Thomson optical depth. In the case the cluster can be assumed
to be isothermal the Comptonization parameter can be expressed by
\begin{equation}
y = \left( {k_{{\rm B}} T_e \over m_e c^2} \right) \tau.
\label{equ:tsziso}
\end{equation}
In this paper we concentrate on the thermal SZ effect and treat the
kinematic merely as a contaminant to the thermal SZ.

\section{Simulating the Sunyaev-Zel'dovich Effect on the full sky}
\label{sec:szsim}

\subsection{The cluster halo distribution}

There are two possible ways to obtain the distribution of cluster
halos in redshift space. The first one is based on assuming a cluster
mass function (see e.g. Press \& Schechter (1974); Sheth \& Tormen
(1999); Jenkins et al.\,(2001); Evrard et al.\,(2002) for cluster mass
functions). This approach
determines the number of clusters within a given mass range at
redshift $z$ per comoving volume element,
\begin{equation}
{{dN}\over{dz}} ~ = ~ \Delta\Omega
{{dV}\over{dzd\Omega}}(z)\int_{M_{\rm lim}}^\infty {{dn(M,z)}\over{dM}}dM,
\end{equation}
where ${{dV}/{(dzd\Omega)}}$ is the comoving volume element and
$({{dn}/{dM}})dM$ is the comoving density of clusters of mass $M$.

\begin{table}
\begin{center}
\begin{tabular}{ccccccc}
Model & $\Omega_0$ & $\Omega_\Lambda$ & $h$ & $\Gamma$ & $\sigma_8$ &
$m/10^{12}\hMsol$\\  \hline $\tau$CDM & 1.0 & 0.0 & 0.5 & 0.21 & 0.6 &
2.22\\ $\Lambda$CDM & 0.3 & 0.7 & 0.7 & 0.21 & 0.9 & 2.25\\
\end{tabular}
\caption{Key parameters of the Hubble Volume simulations. $\Omega_0$
is the density parameter, $\Omega_\Lambda \equiv
\Lambda/3H_0^2$ the cosmological constant, $h \equiv H_0/100 \kmsMpc$ the Hubble constant, $\Gamma$ the power
spectrum shape parameter, $\sigma_8$ the variance of linear
fluctuations smoothed with a top-hat filter on the (comoving) scale of 
8 $\hMpc$ at $z=0$ and $m$ the N-body particle mass.
\label{tab:param}}
\end{center}
\end{table}

\begin{figure}
\begin{center}
\includegraphics[scale=0.42]{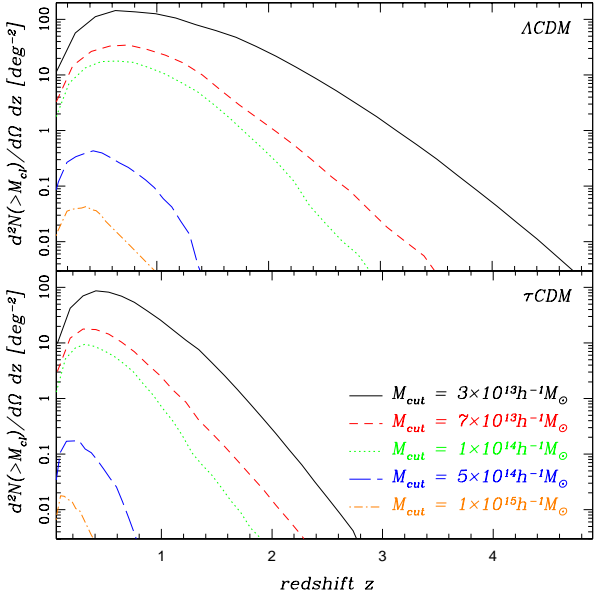}
\caption{The cosmology dependent differential surface cluster number
count per redshift bin $\Delta z=1$ for different lower cluster mass
limits. The upper panel shows the number counts obtained from the
fiducial $\Lambda$CDM simulation and the lower panel the ones for
fiducial $\tau$CDM model (see Table \ref{tab:param}). A typical
cluster detected by Planck has a mass of approximately $5\times
10^{14}\hMsol$. The lower mass clusters provide a realistic
cluster-cluster confusion noise. \label{fig:num_red}}
\end{center}
\end{figure}

The second approach is based on performing N-body simulations and
extracting halos from them to obtain halo catalogues which include
halo masses, three-dimensional positions and the halo
three-dimensional velocities. Such catalogues have been obtained by
applying a spherical overdensity algorithm for the `Hubble volume'
simulations (see Frenk et al.\,2000), the largest N-body simulations
yet performed. Since, besides the halo redshift distribution, N-body
simulations also provide information about halo clustering and
dynamics, we decided to utilise these N-body simulations. These
simulations investigate structure formation in two cosmologies, a
$\tau$CDM and $\Lambda$CDM model. They provide halo catalogues of
lightcone datasets for a
sphere, two octants and a wedge for each cosmological model (see Table
\ref{tab:param}). To obtain a full-sky realisation of the SZ effect we
make use of the sphere and octant datasets. Nevertheless, especially
in the case of the $\Lambda$CDM model, the redshift cut-off of even the
octants ($z\approx 1.5$) is rather low, since objects of intermediate masses ($M \approx
1\times 10^{14}\hMsol$ and below) exist at higher redshift. These objects contribute significantly to
the background cluster confusion, given the $\Lambda$CDM cluster
redshift distribution (see Figure \ref{fig:num_red}). Therefore we
extend the N-body simulations to higher redshifts by using a mass
function to obtain the differential number density,
\begin{eqnarray}
\frac{dn}{d\mbox{ln}M}\left(M,z\right) & = & - A
\frac{\rho_m}{M}\frac{d\mbox{ln}\sigma(M)}{d\mbox{ln}M} \times
\nonumber\\ & &
\exp\left\{-|B-\mbox{ln}\left[g(z)\sigma(M)\right]|^{\epsilon}\right\}\, ,
\label{equ:diffnumdens}
\end{eqnarray}
whose calibration has been  directly derived from the Hubble volume (HV)
simulations (Evrard et al.\,2002) and is thus consistent. These extra
high redshift halos due to the extension of the simulations are mainly
added to provide a realistic SZ background confusion. In equation
(\ref{equ:diffnumdens}), $\rho_m=\Omega_m(0)3H_0^2/(8\pi G)$ is the
present mass density, $\sigma(M)$ the variance of the density field
smoothed with a spherical top-hat filter enclosing mass $M$ at the
mean density and $g(z)$ is the linear growth factor normalised to
$g(0)=1$. In the case of the $\Lambda$CDM model Evrard et al.\,(2002)
find the best fit mass function parameters to the HV simulations to be
$A=0.22$, $B=0.73$ and $\epsilon =3.86$ and for the $\tau$CDM
cosmology $A=0.27$, $B=0.65$ and $\epsilon =3.77$.

\subsection{Halo properties from scaling relations}

Besides the mass of the dark matter halo, which is directly obtained
from the N-body simulations, further cluster properties, such as the
electron gas temperature $T_e$ and virial radius $r_v$, are needed in
order to model the SZ effect of galaxy clusters.

\subsubsection{The mass-temperature relation}
\label{sec:mtrel}

Assuming that clusters are self-similar, virialised and isothermal,
the mass-temperature (M-T) relation can be derived analytically
(Kaiser 1986) and takes the form:
\begin{equation}
\frac{M_{\rm cl}}{10^{15}h^{-1}\mbox{M}_{\odot}}=\left[
\frac{\beta}{(1+z)(\Omega_m
\Delta_c/\Omega_m(z))^{1/3}}\frac{T_e}{\mbox{keV}} \right]^{3/2},
\label{equ:m-t_rel}
\end{equation}
where $M_{\rm cl}$ is the mass of the cluster within its virial radius,
$\Delta_c$ is the mean overdensity inside the virial radius in units
of critical density and  $\beta$ parametrises the scaling relation. In
general, $\beta$ may be a function of $T_e$ as well as cosmology and redshift. In
the case of a flat cosmology ($\Omega_{\Lambda}+\Omega_m=1$)
\begin{equation}
\Delta_c \approx 18\pi^2+82(\Omega_m(z)-1)-39(\Omega_m(z)-1)^2
\end{equation}
(Eke et al.\,1996). While in theory $\beta \approx 0.75$\footnote{Note
that the theoretical prediction is based on a number of assumptions,
such as hydrostatic isothermality of the cluster and $\beta$-profile
adherence of the cluster gas (see e.g. Sarazin
1986).} and is
therefore independent of $T_e$ and cosmology, this result, which has
also been found by numerical simulations (see Pierpaoli et
al.\,2001, for a listing of results), differs from the ones obtained by
observations which favour lower normalisations, i.e. $\beta \approx
0.54(T_{\rm ew}/6$keV$)^{0.24}$, where $T_{\rm ew}$ is the emission-weighted
X-ray temperature of the intracluster gas. This observational
normalisation has been derived from Finoguenov et al.\,(2001) by
converting the mass at a mean overdensity of 500 times the critical
density to the virial cluster mass (see Hu \& Kravtsov 2003).
Note that the findings of Finoguenov et al.\,(2001) are in good
agreement with Allen et al.\,(2001).
To do the conversion, it has been assumed that the dark
matter distribution, which constitutes most of the cluster mass, obeys
a NFW profile, since CDM is the major cluster mass component and
simulations show that its distribution is well described by the NFW
profile (see e.g. Navarro et al. 1997). Although there has been some
controversy, this assumption is supported by observations. For example,
by testing different dark
matter models against mass profiles derived from an XMM-Newton
observation of the relaxed cluster A478 under the hypothesis of
hydrostatic equilibrium, Pointecouteau et al.\,(2004) find that the
NFW profile is significantly preferred to other models. To distribute
intracluster electron gas, however, we use a $\beta$-model. This
assumption is supported by X-ray observations of relaxed galaxy clusters
(see e.g. Ettori et al.\,2004). In the case of isothermality and hydrostatic equilibrium it has
been shown that the expected gas profile of a NFW dark matter halo is well
approximated by such a $\beta$-profile (Eke, Navarro
\& Frenk 1998). Note
further, that in the case of isothermality, $T_e=T_{\rm ew}$, the
observational result predicts a lower mass at fixed $T_e$ than is
predicted by theory. The assumption that clusters are isothermal is
in good agreement with recent XMM-Newton observations of outer cluster
regions which find that the
cluster temperature profiles are isothermal within $\pm 10$ per cent
 up to approximately half the virial radius (Arnaud et al.\,2003).

In the following we will use both normalisations of the M-T relation,
the observed as well as the theoretical one, to analyse the
difference it makes and if a Planck SZ survey will be able to distinguish
between them.

\subsubsection{The integrated Comptonization parameter $Y$ and the total cluster flux density $S_{\nu}$}

After specifying the cluster abundance and the mass-temperature
relation, we relate the observable flux due to the thermal SZ effect
to the cluster mass. Since the spectral behaviour of the thermal SZ
effect is the same for all clusters (see equation \ref{equ:tszspec}),
the total SZ flux density can be expressed in terms of the integrated
Comptonization parameter
\begin{equation}
Y=r_d^{-2} \int y dA,
\label{eq:netot}
\end{equation}
where $d\Omega=dA/r_d^2$, $r_d$ is the angular diameter distance to
the galaxy cluster and one integrates over the cluster's projected
area. Then the total flux density $S_{\nu}$ at frequency $\nu$, which
is usually given in units of Jy, is
\begin{equation}
S_{\nu}(x)=\frac{x^{4}e^{x}}{(e^{x}-1)^{2}} \left[ x
\frac{e^{x}+1}{e^{x}-1} - 4 \right]\,Y=g(x)Y,
\label{eq:totszflux}
\end{equation}
where again $x=h\nu/k_{B}T_{0}$.

In our case, where we assume the cluster to be isothermal, the
integrated Comptonization parameter is further given by
\begin{eqnarray}
Y & = &\frac{k_{B}\sigma_{T}}{m_{e}c^{2}} r_d^{-2} T_{e} \left[\int
 \left(\int n_{e} dl \right)dA \right] \nonumber\\ & = &
 \frac{k_{B}\sigma_{T}}{m_{e}c^{2}} r_d^{-2} T_{e} N_{e},
\label{eq:netot}
\end{eqnarray}
where we have used equation (\ref{equ:tsziso}), and $N_e$ the total number
of thermal electrons within the cluster.  The total number of
electrons within a cluster's virial radius is proportional to the
virial mass of the cluster,
\begin{equation}
N_e=\left( \frac{1+f_H}{2 m_p}\right) f_{gas} M_{cl},
\label{eq:netot}
\end{equation}
where $f_{gas}$ is the baryonic gas mass fraction of the total cluster
mass, $f_H$ is the hydrogen fraction of the baryonic mass -- in the
following we assume $f_H = 0.76$ -- and $m_p$ is the proton
mass. Therefore, the integrated Comptonization parameter $Y$, or
respectively the total flux density, of the cluster can be related to
the cluster's virial mass by
\begin{eqnarray}
Y&=&\frac{k_{B}\sigma_{T}}{m_{e}c^{2}} r_d^{-2} T_{e} \left(
\frac{1+f_H}{2 m_p}\right) f_{gas} M_{cl} \nonumber\\ &=&0.1\times
(1+f_H)\,f_{gas}\,\left(\frac{T_{e}}{keV}\right)\left(\frac{M_{cl}}{10^{15}h^{-1}M_{\odot}}\right)\nonumber\\
&&\times
\left(\frac{r_d}{100h^{-1}\mbox{Mpc}}\right)^{-2}\,h\,\mbox{arcmin}^{-2}
\nonumber\\ 
&=&0.1\times
(1+f_H)\,f_{gas}\,\beta^{-1}(1+z)  \nonumber\\
&&\times \left(\frac{\Delta_c \Omega_m}{\Omega_m(z)}\right)^{1/3}
\left(\frac{M_{cl}}{10^{15}h^{-1}M_{\odot}}\right)^{5/3} \nonumber\\
&&\times 
\left(\frac{r_d}{100h^{-1}\mbox{Mpc}}\right)^{-2}\,h\,\mbox{arcmin}^{-2}.
\label{eq:Y-M_rel}
\end{eqnarray}
Hence, apart from the normalisation, $Y$ depends on the cluster
gas mass, the cosmology and the cluster redshift.

\subsubsection{The cluster electron gas density profile}

Since the $\beta$-model has been empirically found to fit well the electron
gas density profile obtained from X-ray observations, and since
most of the clusters will be unresolved given the resolutions of the
Planck channels (see section \ref{sec:planckobssim}), we utilise a spherical isothermal $\beta$-profile,
\begin{equation}
\rho(r)=\rho_{0}\left[1+\left(\frac{r}{r_c}\right)^2\right]^{-3/2\beta_{\rm prof}},
\label{eq:betaprof}
\end{equation}
to model the electron gas distribution inside the cluster's virial
radius. The line-of-sight projected $\beta$-profile is given
by:
\begin{equation}
n_e^{\rm proj}(r)=n_{e0}\left[1+\left(\frac{r}{r_c}\right)^2\right]^{1/2-3/2\beta_{\rm prof}},
\label{equ:projpro}
\end{equation}
where $r_c$ is the cluster core radius and
$n_{e0}$ is the central electron gas density. For our simulations we
make the common choice of $\beta_{\rm prof} = 2/3$.
While this choice of $\beta$ is consistent with the
assumption of hydrostatic equilibrium for the virial M-T relation
normalisation, for the one derived from X-ray observations the
resulting cluster $y$-profile given the gas density mimics a
temperature gradient. When a truncation of the
cluster electron gas profile (\ref{eq:betaprof}) is performed at the virial radius, the projected profile
becomes
\begin{equation}
n_e^{\rm proj}(r) = \frac{2\,n_{e0}}{\sqrt{1+{\left( r/r_{c}
\right)}^{2}}} \times \mbox{arctan} \sqrt{\frac{{b}^{-2}-{\left(
r/r_{c} \right)}^{2}}{1+{\left( r/r_{c} \right)}^{2}}},
\label{equ:truncprojpro}
\end{equation}
where $b$ relates the cluster core radius to the virial radius (see below).
This truncation ensures that the cluster SZ signal drops to zero
within a finite range from the cluster centre. Thus, in the cluster
outskirts the projected profile (\ref{equ:truncprojpro}) declines more
rapidly than the one given by (\ref{equ:projpro}). The resulting
change in the shape of the projected profile also agrees better
with predictions obtained from hydrodynamical simulations.

As mentioned above, we use the virial radius to define the
edge of the cluster. Since $M_{cl}=\frac{4}{3}\pi \bar{\rho}
r_v^3$, where $\bar{\rho}$ is the mean matter density inside $r_v$,
the virial radius of the cluster is given by:
\begin{eqnarray}
r_v&=&2.12\times \left(
\frac{M_{cl}}{10^{15}h^{-1}M_{\odot}}\right)^{1/3}(1+z)^{-1} \nonumber\\
&&\times \left( \frac{\Omega_m \Delta_c}{\Omega_m(z)}
\right)^{-1/3}\,h^{-1}Mpc.
\label{equ:radvir}
\end{eqnarray}
Moreover, we relate the cluster core radius $r_c$ to its virial radius
$r_v$ by $r_c(z) = b(z)\,r_v$, where $b(z=0)\approx 0.14$ and the
evolution of the cluster profile can be described by:
\begin{equation}
r_c(z)=r_c(z=0)\times (1+z)^{1/5}.
\end{equation}
The scaling $b(z)$ has been chosen so that the evolution of the
central electron density of the cluster $n_{e0}$ is consistent with
simulations (see e.g. Eke et al.\,(1998)).

Given the shape of the electron gas density profile, the electron gas
distribution and therefore the optical depth can be normalised in
terms of the cluster's virial mass by utilising equations
(\ref{eq:netot}) and (\ref{eq:betaprof}) and by making use of
\begin{equation}
\int^{r_v}_{0} \rho_{gas}(r)4\pi r^2\,dr=f_{gas}M_{cl}.
\label{eq:rv}
\end{equation}
Hence, one deduces that the central electron gas density is given by:
\begin{eqnarray}
n_{e0}&=&1.6\times10^{3}h^2f_{gas}\left(\frac{1+f_H}{w}\right)\left(\frac{r_c}{h^{-1}\mbox{Mpc}}\right)^{-3}
\nonumber\\
&&\left(\frac{M_{cl}}{10^{15}h^{-1}M_{\odot}}\right)\mbox{m}^{-3},
\end{eqnarray}
where $w$ is given by
\begin{equation}
w=b^{-1}+\mbox{arctan}\left(b\right)-\frac{\pi}{2}
\end{equation}
when we assume a $\beta$-profile with $\beta_{\rm prof}=2/3$. Note that --
besides the relation for the central electron gas density -- all
scaling relations between cluster properties, which are given above,
are independent of the chosen electron gas profile.

Since the scaling relations presented above are for virial cluster
properties, we have to rescale the cluster masses given by the Hubble
volume cluster catalogues and the mass function of Evrard et
al.\,(2002), as they provide the cluster mass within a radius for which
the mean density contrast $\Delta_c = 200$. The cluster mass
rescaling is discussed in section \ref{sec:convclm}.
 
\subsection{Varying $\sigma_8$}

In order to rescale the cluster halo dark matter masses to values of
$\sigma_8$ which differ from the default $\sigma_8$ used in the
conducted HV N-body simulations we follow the formalism presented in
Evrard et al.\,(2002). A relation between the change in $\sigma_8$,
$\Delta \sigma_8$, and the related change in cluster mass, $\Delta
M$, can be derived by setting to zero the total derivative of the
differential comoving cluster number density (equation
\ref{equ:diffnumdens}) at fixed redshift $z$. The resulting relation
is given by
\begin{equation}
\frac{\Delta \sigma_8}{\sigma_8} \ = \alpha^\prime(M)  \ \frac{\Delta
M}{M}
\label{eq:dlnsig8dlnM}
\end{equation}
with
\begin{eqnarray}
\alpha^\prime(M) & = & \alpha_{\rm eff}(M)+ \frac{(1-2b/\alpha_{\rm eff}(M))}{\epsilon \,
(\lnsiginv (M) + B)^{\epsilon-1}}
\label{eq:alphaprime}
\end{eqnarray}
where $\alpha_{\rm eff}(M) \equiv \dlnsiginv(M)/\dlnM$ and can be
obtained by employing the quadratic relation describing the filtered
power spectrum shape given in Evrard et al.\,(2002). As has been
pointed out by Evrard et al.\,(2002), the number of the rarest most
massive clusters is most sensitive to a change in $\sigma_8$.

This expression (equation \ref{eq:alphaprime}) differs from the one in Evrard et al.\,(2002), since
we do not make any approximation.
In particular, for lower masses this expression is slightly more
accurate (10 per cent for $M_{cl} = 1\times 10^{14}\hMsol$).
The presented rescaling relation between the halo
dark matter mass and the change in $\sigma_8$ has been used to convert
the CDM cluster masses of the fiducial HV simulations (see Table
\ref{tab:param}) to the ones of models with different values of
$\sigma_8$, a still rather unconstrained parameter by today's
observation.

\subsection{Converting to virial mass}
\label{sec:convclm}

Commonly, a cluster halo is defined to be a spherical overdense region
- in the case of N-body simulations - of particles, whose internal
mean density, $\langle\rho\rangle$, is given as a multiple either of the
critical density $\rho_c(z)=3\,H(z)^2/8\pi\,G$ or of the overall mean
matter density $\bar{\rho}(z)=\Omega_m(z)\,\rho_c(z)$. For the HV
catalogues $\Delta_c=200$, where
$\langle\rho\rangle=\Delta_c\rho_c(z)$. In the case of the $\tau$CDM
model this is not far-off the prediction of $\Delta_c \approx 178$
given by the spherical top-hat collapse model for a sphere that
encloses the virial mass, $M_{cl}$ (see Lacey \& Cole (1993)). For
the fiducial $\Lambda$CDM cosmology however this value differs approximately
by a factor of two from the predicted collapse model value of
$\Delta_c(0)\approx 103$ at redshift $z=0$, and $\Delta_{\bar{\rho}}(0)
\approx 667$ for the HV simulations. Since the scaling relations presented above, especially
the M-T relation (see equation \ref{equ:m-t_rel}), are derived for the
virial cluster properties, either the cluster mass has to be directly
converted to yield the virial mass of the cluster or one has to
rescale the relations by deriving an approximate formula to the
cluster mass conversion. Note, in either case a halo density profile
has to be assumed in order to be able to convert various definitions
of the cluster mass into each other.

Therefore, in this paper we proceed by assuming a NFW halo profile of
the dark matter which contributes mainly to the total cluster mass and
follow the recipe as it has been suggested in Hu \&
Kravtsov (2003) to rescale the dark matter halo masses.

\subsection{Simulating Planck Observations}

\begin{table*}
\begin{center}
\begin{tabular}{||l||r|r|r|r|r|r|r|r|r|r||} \hline
Instrument & \multicolumn{3}{c|}{LFI}& & \multicolumn{6}{c||}{HFI}\\
\hline
Central frequency (GHz) & 30 & 44 & 70 && 100 & 143 & 217 &
353 & 545 & 857\\
\hline
Angular resolution & 33 & 24 & 14 && 9.2 &
7.1 & 5 & 5 & 5 & 5\\
(FWHM, arcmin) & & & & & & & & & & \\
\hline
Sensitivity per pixel ($\mu K$) & 4.5 & 6 & 10.5 && 4.5 & 4.9 & 10.7 & 32.7 & 327.7 & 14935 \\
(after three full sky coverages) & & & & & & & & & & \\
\hline
\end{tabular}
\caption{Basic observational characteristics of the $9$ frequency
channels of the Planck instruments (Low Frequency Instrument (LFI) and High Frequency Instrument (HFI)). The angular resolutions are quoted as FWHM for a Gaussian beam. The sensitivities quoted will be achieved by Planck after three full sky coverages. A pixel is a square whose side is the FWHM extent of the beam. \label{planckinst}}
\end{center}
\end{table*}

The thermal and kinematic SZ effect contributions to a full-sky map
at each Planck channel are obtained from catalogues of
cluster properties by employing the relations and profile discussed
above. Further components regarded as
contamination of the clusters' SZ signal were also simulated. The
following paragraphs contain short
descriptions of the realisation of these further components.

\subsubsection{Primordial CMB}

The realizations of full-sky maps of the
primordial CMB for different cosmological models are generated by
assuming that the primordial CMB is a
homogeneous random Gaussian field entirely described by a standard CDM
power spectrum whose shape and normalisation agrees with WMAP measurements (Bennett et al.\,2003). The $C_l$ coefficients have been created using CAMB
(Lewis, Challinor \& Lasenby 2000).

\subsubsection{Point sources}

Point source contamination has been included in our simulations. Here
we follow the modelling approach presented in Hobson et
al.\,(1999). In respect of the component reconstruction, which is
performed in spherical harmonics space (see section \ref{sec:hsmem}), we make the
assumption that the point source contribution in each Planck
frequency channel should be well modelled as random
Gaussian noise in the Fourier or spherical harmonics domain. The
contribution by the point source population is
determined by number counts predicted by Toffolatti et al. (1998)
based on the evolution model of radio sources of Danese et al. (1987)
and the far-IR source evolution model by Franceschini et al. (1994),
which has been updated by Burigana et al. (1997) to account for an
isotropic sub-mm component. The rms thermodynamic temperature
fluctuation within a defined pixel area due to the point source
population in each Planck channel after beam-convolution is shown in
Figure 4 of Hobson et al. (1999) for sources in the
$10^{-5}$Jy$<S(\nu)< 10$Jy flux range. We
assume no correlation of the point sources with each other or with
clusters of galaxies. In the case of radio point sources however it is
known that they have in fact a correlation with galaxy clusters. Thus, our
assumption leads to some underestimate of confusion at the lowest
Planck frequencies.
If IR sources originate at high redshift
($z>3$) as suggested by current models as well as SCUBA observations
at 850-$\mu$m (Holland et al.\,1998; Smail et al.\,1998; Blain et
al.\,1998), the assumption of IR sources being uncorrelated with
clusters which are detectable by Planck (see section \ref{sec:planckclnum}) is
valid. Generally, the contribution of clustered sources is found to be
small in comparison to the Poisson term. Note that the Toffolatti
model has been found to agree within a factor of two with results obtained from the WMAP
mission (Bennet et al.\,2003). The WMAP results suggest that the model
actually overpredicts the point source contribution. This is also
supported by recent VSA 9C survey counts (see Figure 13 in
Bennett et al.\,2003). Thus, our modelling is rather pessimistic concerning
the radio point source contribution. IR sources are expected to
dominate the number counts of bright objects at frequencies $\nu >
100$ GHz, whereas radio sources are a significant contaminant below
$\nu = 100$ GHz. Currently
the best observational limits on the IR source contribution are
obtained from SCUBA. These observations suggest that the IR source
contribution could be up to a few times higher than predicted by the
Toffollati model at 353 GHz. However, since the SCUBA surveyed area is
rather small (a few 10 arcmin$^{2}$) and IR sources
relevant to Planck are at the bright end of the SCUBA population, the sample
variance is quite large and does not yield any final conclusions about
the IR source contribution.

\subsubsection{Galactic dust emission}

The dust emission is modelled using the DIRBE-IRAS 100-$\mu$m dust map
(Schlegel, Finkbeiner \& Davis 1998). Its
resolution of 5 arcmin is sufficient for our modelling purpose since it
is comparable to the high frequency channel resolutions of Planck. To
extrapolate the flux at 100-$\mu$m to the Planck observing frequencies
a one-component dust model with a temperature of $T_{dust}=18$K and a
dust emissivity $\beta =2$ is assumed by taking into account the
colour correction of the DIRBE 100-$\mu$m filter.

\subsubsection{Galactic synchrotron emission}

The destriped version of the 408MHz Haslam survey (Haslam et
al.\,1982), to which artificial substructure on sub-degree scale has
been added by extrapolating the power spectrum, is used as the synchrotron
emission template. Its extrapolation to 300 GHz has been obtained by
using an all-sky spectral index map derived from low frequency surveys
at 408 MHz, 1420 MHz (Reich \& Reich 1986) and 2326 MHz (Jonas, Baart \&
Nicholson 1998) and padding the unobserved area around the South pole
with the mean spectral index. By assuming a constant spectral index
$\beta = 0.9$ the synchrotron emission at the Planck channel
frequencies has been predicted.
  
\subsubsection{Galactic free-free emission}

A full-sky free-free emission template has been created on the basis
of the DIRBE/IRAS dust map. 60 per cent of the free-free emission is
assumed to be dust correlated. The dust uncorrelated free-free
component has been obtained by flipping the dust map north-south in
Galactic coordinates. A spectral index of $\beta=-0.16$ has been
assumed and the normalisation has been taken from Bouchet \& Gispert
(1999).
  
\subsubsection{Planck channel observations}
\label{sec:planckobssim}

Realistic Planck surveyor simulations are then obtained at each of the
$n_f = 9$ observing frequencies (see Table \ref{planckinst}) along the
line-of-sight $\hvect{x}$ by summing up the contributions of the
physical components along this direction, convolving the resulting
anisotropy map with the Planck beam at each particular frequency and
adding the channel noise, which is assumed to be uncorrelated and
Gaussian. This can be conveniently represented as a data vector of
length $n_f$ containing the resulting observing
signal in direction $\hvect{x}$ at each frequency:
\begin{equation}
d_\nu(\hvect{x}) = \int_{4\pi} B_\nu(\hvect{x}\cdot\hvect{x}') 
\sum_{p=1}^{n_c} F_{\nu p}\,s_p(\hvect{x}') \,{\rm d}\Omega'
+ \epsilon_\nu(\hvect{x}),
\end{equation}
where $F_{\nu p} = \int_0^\infty t_\nu(\nu') f_p(\nu') \,{\rm d}\nu'$
is the frequency response matrix, $B_\nu$ is
the beam profile of the $\nu$th frequency channel. Up-to-date beam sizes and
instrumental channel noises for Planck are given on the Planck
homepage\footnote{\it http://astro.esa.int/Planck/science/performance/perf\_top.html}.

In order to reduce the instrumental noise contribution in the
high-frequency Planck channels to a level comparable to the point
source noise, the presented results have been obtained by simulating
Planck observations of one and a half year duration (approximately
three full sky coverages).

Even more realistically one expects anisotropic instrumental noise and
spatial variations of the Galactic dust temperature and spectral
index. Stolyarov et al.\,(2004) have
recently extended their modelling to accommodate these effects and
applied the HSMEM to the simulated Planck channel observations. Since
they find that these spatial variations do not affect the
recovery quality of the components, our modelling is
sufficient to obtain reliable results. In any case, these variations happen on scales
larger than the typical cluster scale and thus the assumption of
locally constant noise and foreground properties in cluster-sized regions is a reasonable one. 

\section{Reconstructing cluster's SZ signal}
\label{sec:recon}

\subsection{The HSMEM}
\label{sec:hsmem}

In order to extract the thermal SZ signal from other CMB components
on the sky, we use the harmonic-space maximum entropy method (HSMEM)
as presented in Stolyarov et al.\,(2002). This method allows us to
perform a full-sky separation of all input components described above
in spherical harmonic space. While the HSMEM has been
developed mainly focusing on the recovery of the primordial CMB
component and its power spectrum, it has been applied in this work to
extract the thermal SZ component and all other components have been
considered as contaminating noise. In the following a short
description of the method is presented. A detailed
explanation of the method can be found in
Stolyarov et al.\,(2002) and Hobson et al.\,(1998).
 
The HSMEM is a Bayesian method. In the spherical harmonic
representation when we neglect any
coupling between different ($l$,$m$) modes, the posterior
distribution, which we want to explore, is given by
\begin{equation}
\Pr({\mathbfss a_{\ell m}}|{\mathbfss D_{\ell m}}) \propto
\Pr({\mathbfss D_{\ell m}}|{\mathbfss a_{\ell m}})\Pr({\mathbfss
a_{\ell m}}),
\end{equation}
where $\Pr({\mathbfss D_{\ell m}}|{\mathbfss a_{\ell m}})$ is the
likelihood of the signal vector ${\mathbfss a_{\ell m}}$ given the
data ${\mathbfss D_{\ell m}}$. The prior probability $\Pr({\mathbfss
a_{\ell m}})$ contains our previous knowledge or assumptions about the
signal vector before acquiring any data. In the case of the
maximum-entropy method the specific feature that
distinguishes it from other Bayesian component
separation algorithms is that it assumes the prior to be of entropic
form, $\Pr({\mathbfss h}_{\ell m}) \propto \exp[\alpha S({\mathbfss
h}_{\ell m},{\mathbfss m})]$, where the `hidden' vector ${\mathbfss
h}_{\ell m}$ is related to the signal vector by ${\mathbfss a}_{\ell
m}={\mathbfss L}_\ell {\mathbfss h}_{\ell m}$ and ${\mathbfss m}$ is a
model vector to which ${\mathbfss h}_{\ell m}$ defaults in the absence
of data. $S({\mathbfss h}_{\ell m},{\mathbfss m})$, whose form has
been derived in Gull \& Skilling (1990) and Hobson \& Lasenby (1998), is
the cross-entropy of ${\mathbfss h}_{\ell m}$ and ${\mathbfss m}$ and
$\alpha$ is a regularisation parameter, which is treated as another
optimisation hypothesis parameter.  ${\mathbfss L}_\ell$ is obtained
by a Cholesky decomposition of the isotropic signal covariance matrix
${\mathbfss C}_{\ell} = \langle {\mathbfss a}_{\ell m} {\mathbfss
a}^\dagger_{\ell' m'} \rangle$. Using such an entropic prior instead
of a Gaussian, as is the case when Wiener filtering, ensures a
better reconstruction of non-Gaussian components such as the thermal
and kinematic SZ effect.

A maximisation of the posterior with respect to ${\mathbfss a}_{\ell
m}$ is therefore equivalent to minimising the function
\begin{equation}
\Phi({\mathbfss h}_{\ell m})=\chi^2({\mathbfss h}_{\ell m}) - \alpha
S({\mathbfss h}_{\ell m},{\mathbfss m}),
\label{function}
\end{equation}
since assuming Gaussian instrumental noise
\begin{equation}
\Pr({\mathbfss d}_{\ell m}|{\mathbfss h}_{\ell m}) \propto \exp
\left[-\chi^2({\mathbfss h}_{\ell m})\right],
\end{equation}
where $\chi^2$ is the standard misfit statistic which utilises the
isotropic noise covariance matrix ${\mathbfss N}_{\ell} = \langle
\bmath{\epsilon}_{\ell m} \bmath{\epsilon}^\dagger_{\ell' m'} \rangle$.

To recover the SZ signal with the HSMEM, prior knowledge about
cross-correlation between different components has not been
assumed. Nevertheless, prior knowledge about the approximate shape and
amplitude of the component power spectra is included in the analysis. This assumption is reasonable since
preceding observations and data analysis of Planck observations
themselves will provide prior
constraints on the component power spectra.

\subsection{Cluster detection algorithm}
\label{sec:cldetec}
 
\begin{figure*}
\begin{center}
\subfigure[$\sigma_8=0.9$
(virial)]{\label{fig:nonres:a}\includegraphics[angle=-90,width=0.3\textwidth]{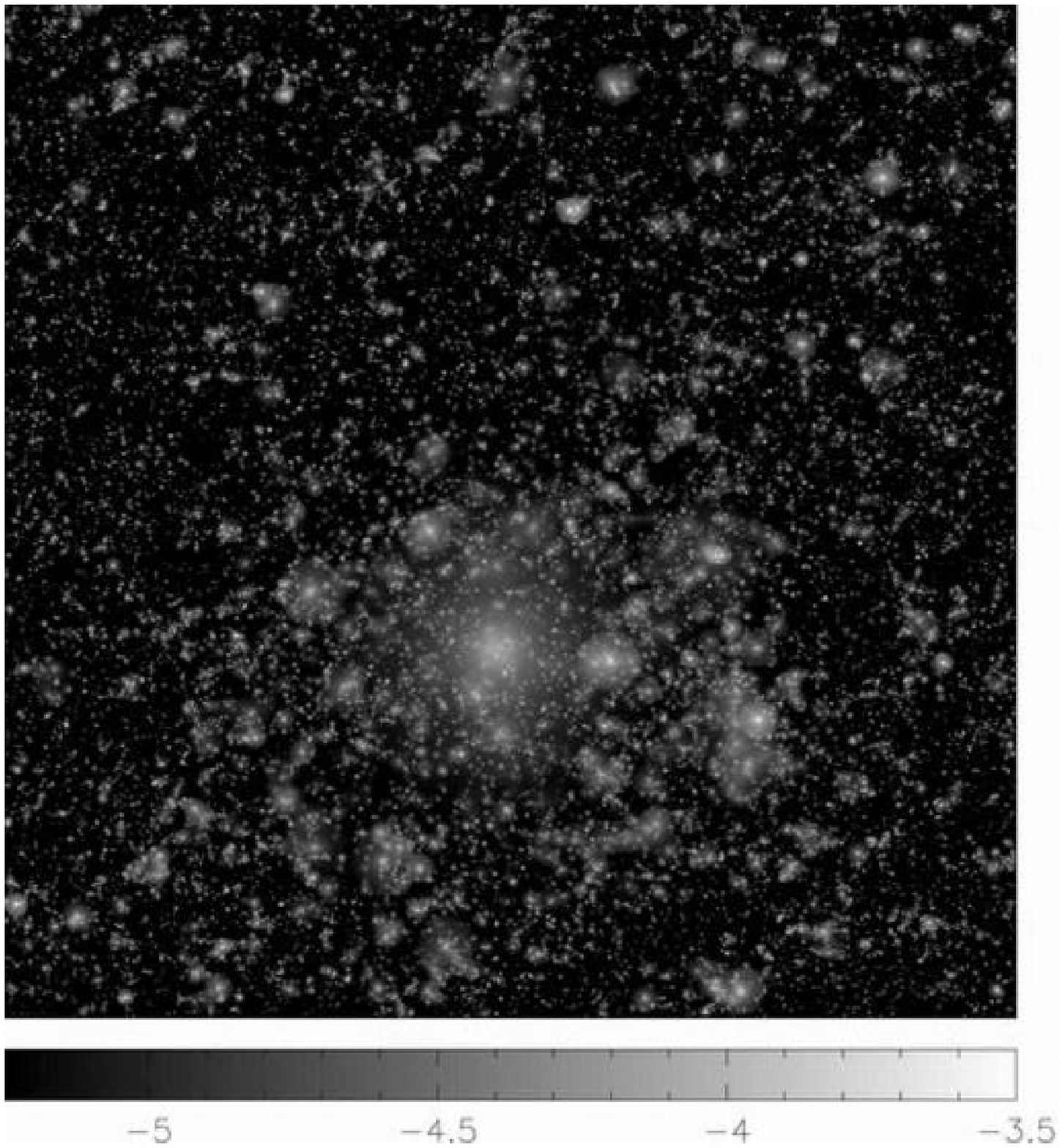}}
\subfigure[$\sigma_8=0.7$
(virial)]{\label{fig:nonres:b}\includegraphics[angle=-90,width=0.3\textwidth]{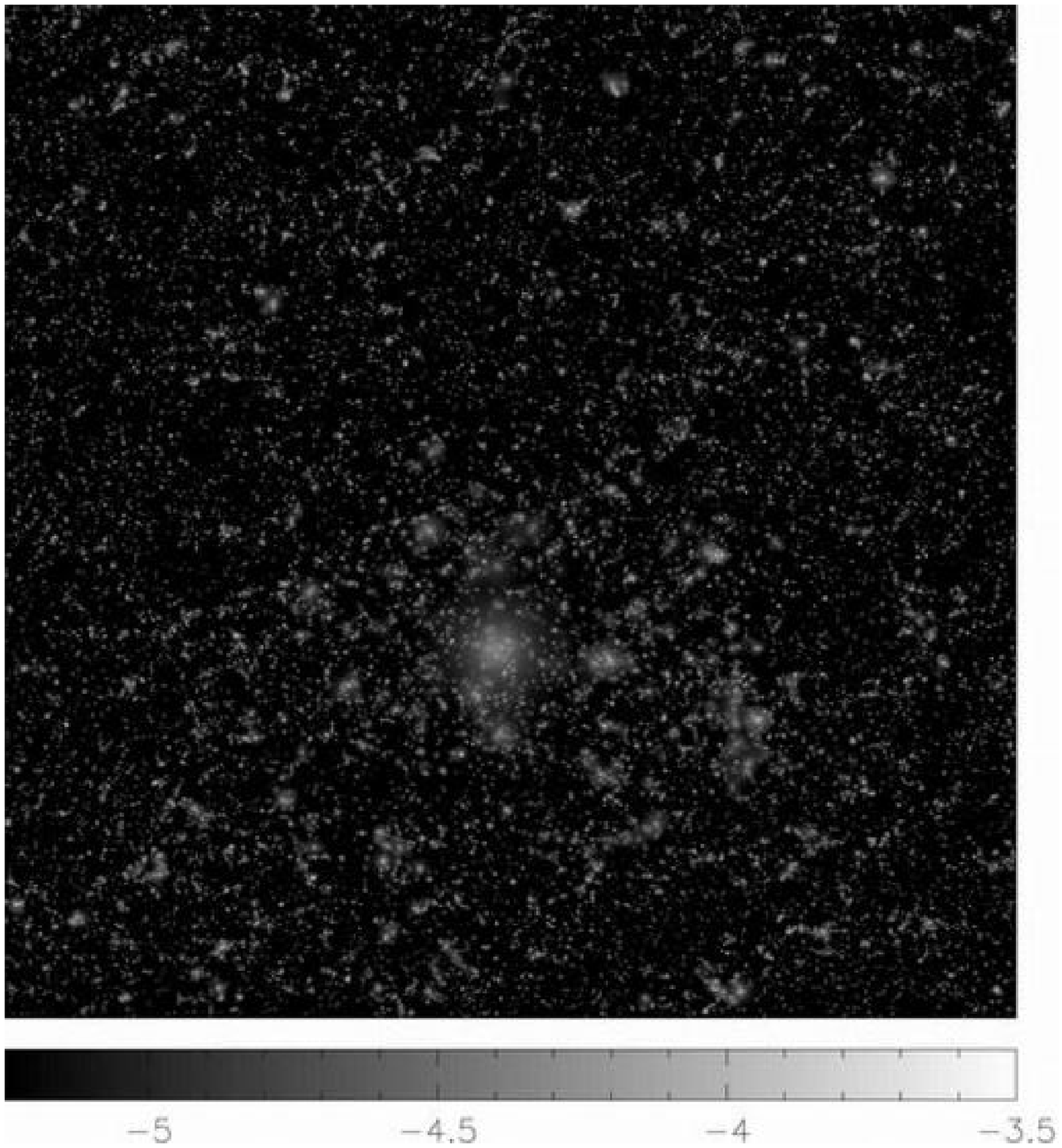}}
\subfigure[$\sigma_8=1.0$
(virial)]{\label{fig:nonres:c}\includegraphics[angle=-90,width=0.3\textwidth]{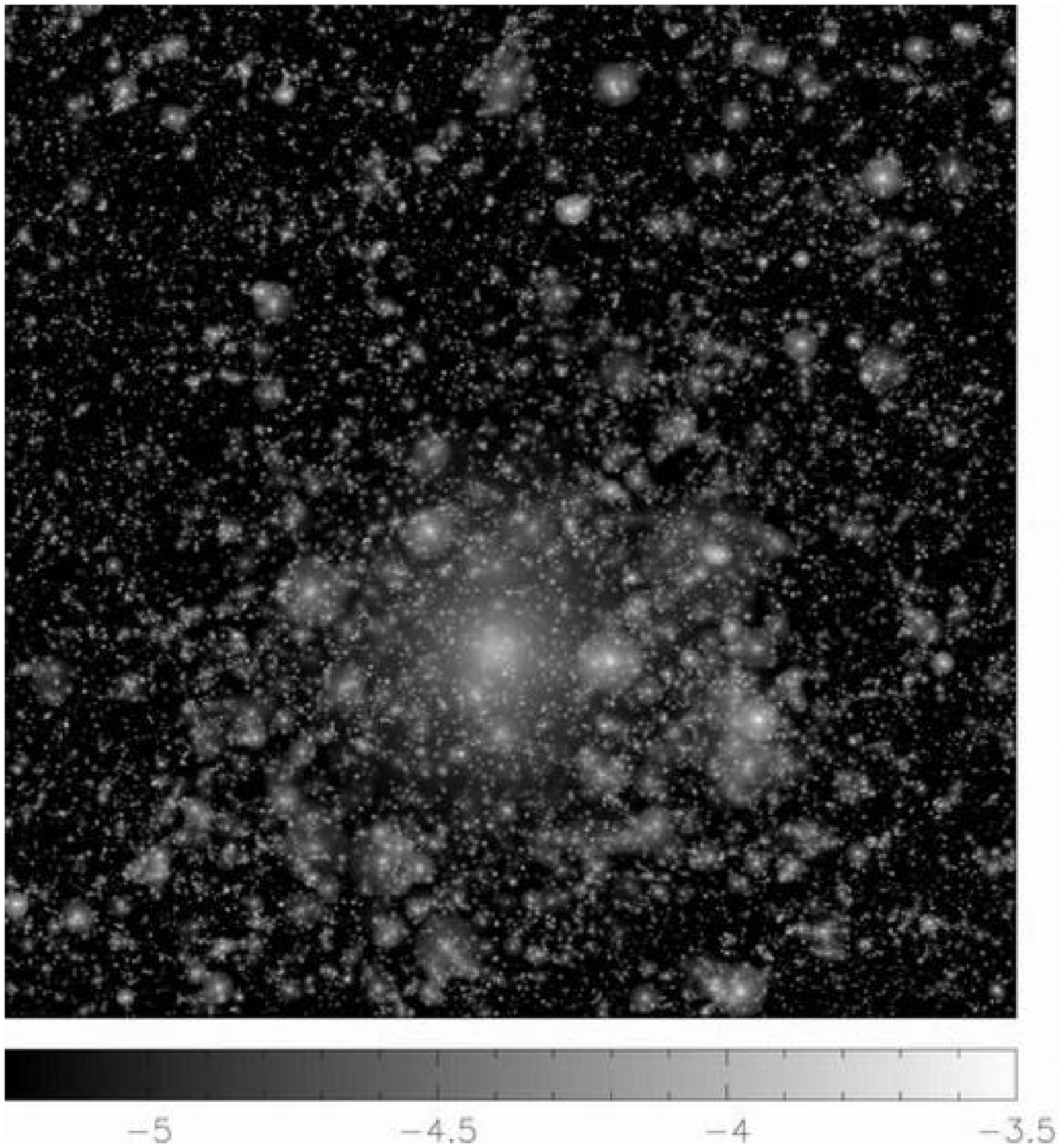}}
\subfigure[$\sigma_8=0.9$
(xnorm)]{\label{fig:nonres:d}\includegraphics[angle=-90,width=0.3\textwidth]{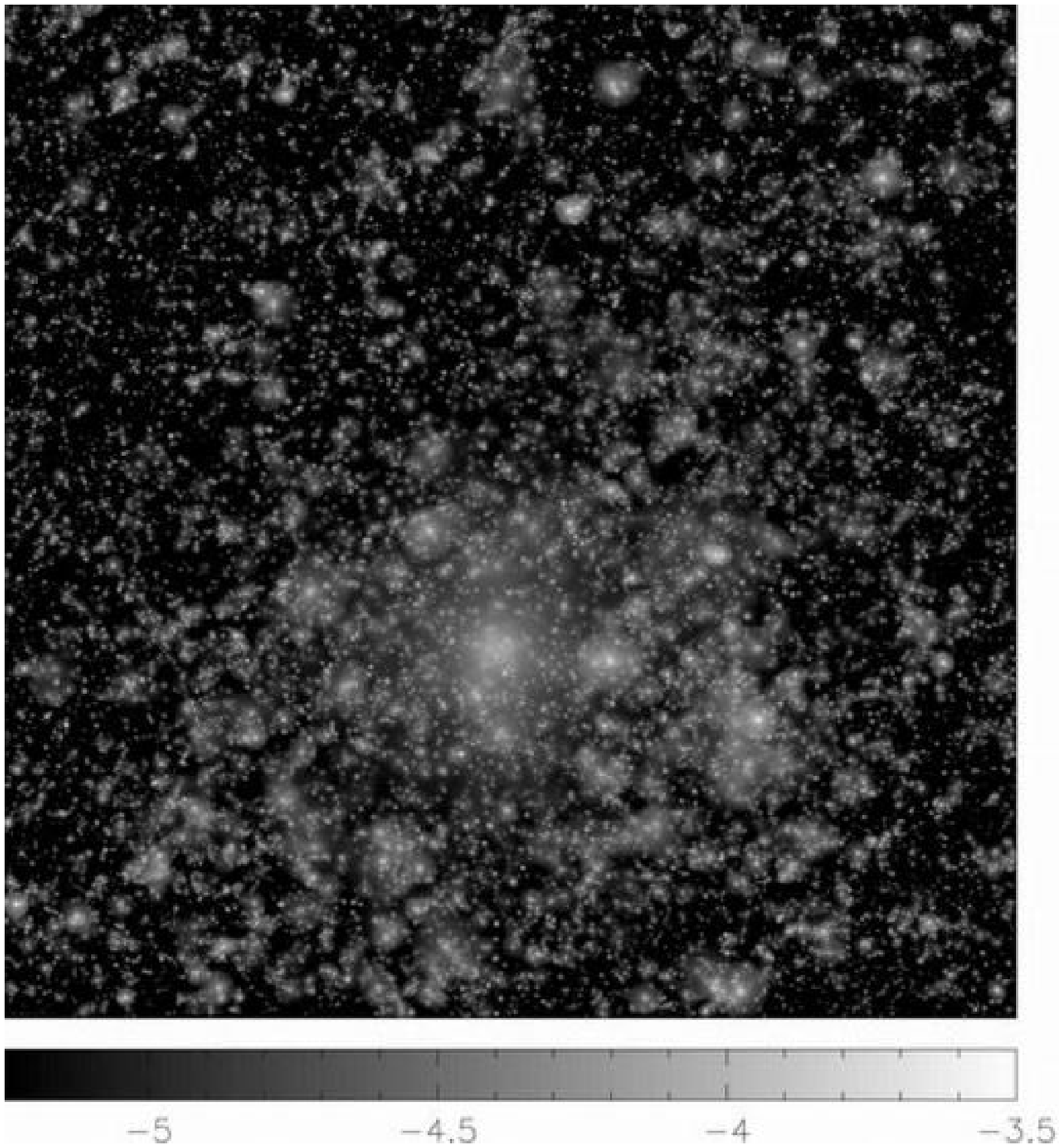}}
\subfigure[$\sigma_8=0.7$
(xnorm)]{\label{fig:nonres:e}\includegraphics[angle=-90,width=0.3\textwidth]{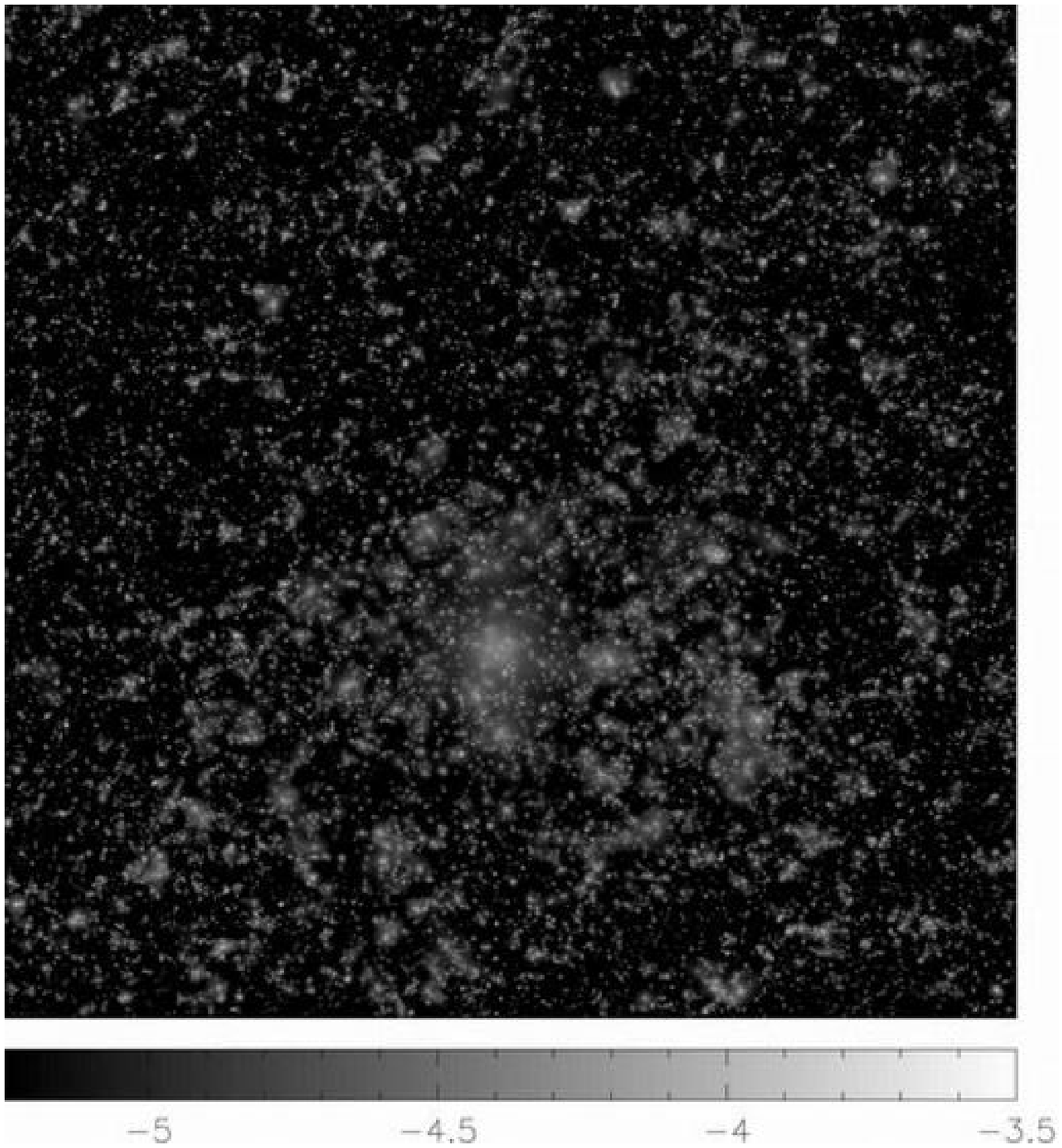}}
\subfigure[$\sigma_8=1.0$
(xnorm)]{\label{fig:nonres:e}\includegraphics[angle=-90,width=0.3\textwidth]{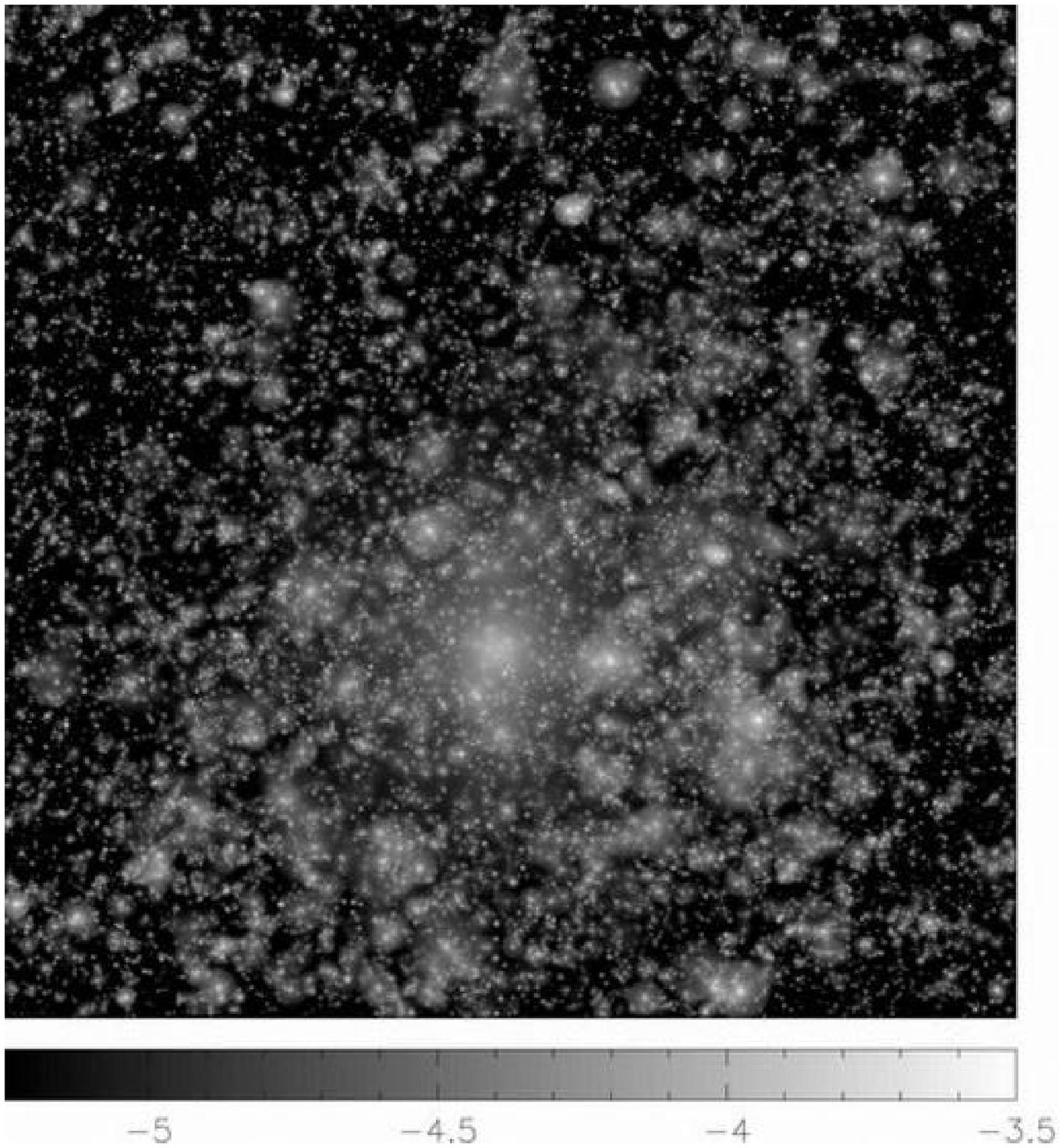}}
\caption{$12.5\times 12.5$deg$^2$ thermal SZ realizations of the same
patch of sky for three different values of $\sigma_8$ and two
different M-T scaling relations (see equation \ref{equ:m-t_rel} and text) for the
$\Lambda$CDM model (see Table \ref{tab:param}) used to make simulated {\it Planck
Surveyor} observations. The value of $\sigma_8$ and the M-T relation
used in the particular realization are given below each panel. The
maps are shown in units of the frequency independent Compton $y$
parameter. \label{fig:tszlcdmrel}}
\end{center}
\end{figure*}

\begin{figure*}
\begin{center}
\begin{minipage}[c]{0.5\textwidth}
\begin{center}
\subfigure[$\sigma_8=0.6$
(virial)]{\label{fig:nonres:a}\includegraphics[angle=-90,width=0.6\textwidth]{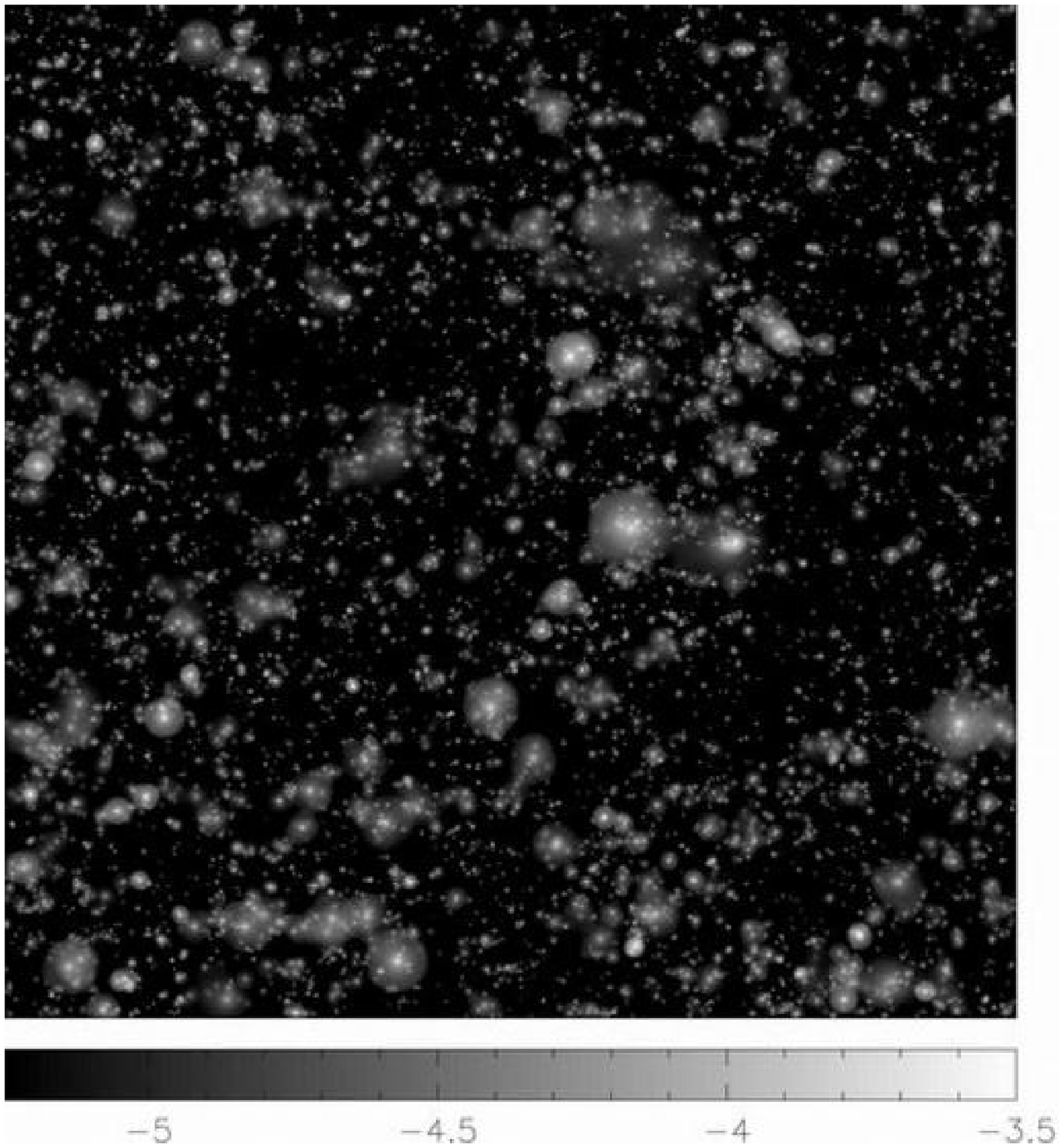}}
\end{center}
\end{minipage}%
\begin{minipage}[c]{0.5\textwidth}
\begin{center}
\subfigure[$\sigma_8=0.6$
(xnorm)]{\label{fig:nonres:c}\includegraphics[angle=-90,width=0.6\textwidth]{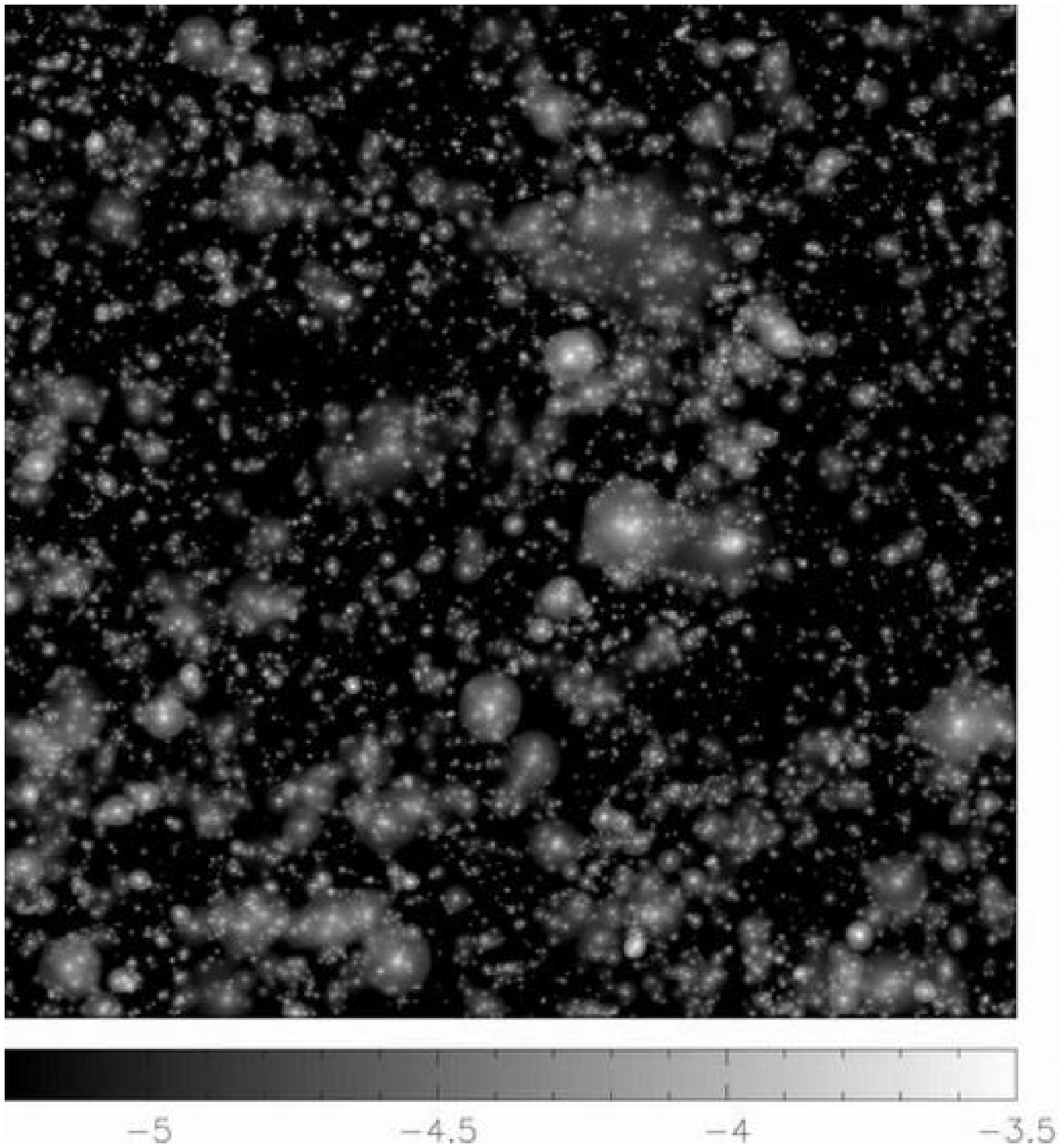}}
\end{center}
\end{minipage}
\caption{The same as Figure \ref{fig:tszlcdmrel} for the $\tau$CDM
model (see Table \ref{tab:param}). Here a value of $\sigma_8 = 0.6$
has been used together with the different M-T relation normalisations to produce the thermal SZ simulations. \label{fig:tsztcdmrel}}
\end{center}
\end{figure*}

To obtain the thermal SZ signal of single clusters the HSMEM recovered
thermal SZ map is searched for local maxima above a certain threshold,
which can be optimised to maximise the number of detected clusters
and to minimise the number of false detections. A cluster is
considered to be detected when the true position of its peak signal coincides with a local
maximum found in the reconstructed map to within a defined off-set around
the peak. In the work presented here an off-set of one map pixel size
($\approx 2$ arcmin; chosen to be just below the Nyquist sampling for
the smallest Planck beam) has been allowed in the case the cluster's
core radius angular extent is smaller than the size of a pixel. This
applies to the vast majority of galaxy clusters. Otherwise the
local maximum (in the following also referred to as SZ
peak detection) has to lie
within the core radius of the true cluster to be
matched. This choice of off-set has been found to be optimal in order to obtain a high matching rate and avoid ambiguities arising from several maxima or clusters being matched to one of the other kind. Only maxima whose amplitude is greater than a fixed
multiple of the estimated noise rms of the reconstructed map,
$\sigma_{noise}$, (minimally $3\times \sigma_{noise}$) have been taken
into account. The rms of the reconstructed map is obtained iteratively
by masking maxima and their surroundings which are likely to belong to
a cluster detection. Starting with the rms obtained over the whole map
we exclude maxima above a threshold, for example $5\times
\sigma_{noise}$, and their environment which is assigned to them by
flux integration. This process is repeated until the rms has converged.

If a maximum can be matched to several clusters or several maxima can
be matched to one cluster then an amplitude and offset comparison is carried
out in order to obtain the most likely match. In the case of
extended clusters it can happen that due to noise (and background
confusion) several maxima belong to the same cluster. Therefore, if
maxima fall within each others estimated radius they are assumed to
relate to the same extended cluster.

The integrated Comptonization parameter $Y$ and cluster radius
$r_{cl}$ associated with a SZ peak detection are estimated by spherical flux
integration. At the radial extent at which the change in the interior
integrated flux falls below a certain threshold, i.e. the
contribution which is expected to be due to $3\times \sigma_{noise}$,
the cluster is cut off. This ensures that the signal-to-noise ratio of
cluster detections is at least 3.

Even though recently more sophisticated algorithms have been
suggested, such as the pseudo optimal filter
(Herranz et al. 2002) and MCMC methods (Hobson \& McLachlan 2003), we
decided to use the presented
algorithm since it is efficient and has the ability to provide robust
results (see section \ref{sec:planckclnum}). The advantage of the
algorithm presented here is
that, besides the assumption of the cluster to be spherical, the flux
estimate is independent of the chosen cluster profile. This is not the
case for a pseudo optimal matched
filter or MCMC application. The assumption that the cluster shape is
spherical should be a fair approximation for most
clusters, since X-ray observations show that the clusters' median
ellipticity is $\approx 0.2$ (see Mohr et al. 1995).

We tested the assumptions on which the matching algorithm relies, such
as the choice of the acceptance region and the association of multiple
SZ peaks to a single extended cluster, and found, even though further
optimisation may still be possible, that our choices are reasonable
and do not affect the obtained completeness and purity estimates to a
great extent. Nevertheless, problems which we found are discussed in the
text below (see section \ref{sec:planckclnum}). Furthermore, we note that it is most likely
that the final analysis performed on the Planck data will consist of a combination of several
methods making optimal use of their different advantages, such as
speed, detection efficiency and prior knowledge, at different
levels of the analysis to maximise the number of cluster
detections and the speed of the analysis.

\section{Results and discussion}
\label{sec:resanddis}

\begin{figure}
\begin{center}
\includegraphics[scale=0.42]{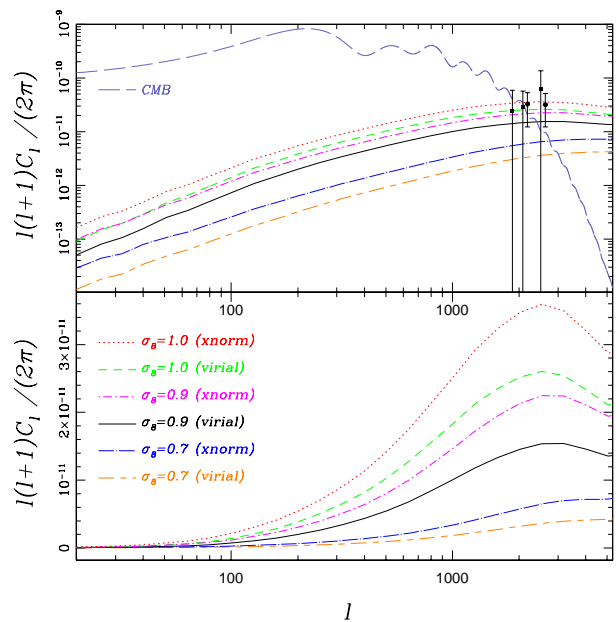}
\caption{The power spectra of the primordial CMB and the SZ effect for
different values of $\sigma_8$ and M-T relations (see panel and text)
at 300 GHz for the $\Lambda$CDM simulations. While the upper panel
shows a log-log plot, in the lower panel the SZ power spectra
amplitudes are plotted on a linear scale in order to
emphasise the different slopes and multipoles at
which the SZ power spectra peak given different values of
$\sigma_8$. The upper panel includes the CMB excess power spectrum
amplitudes as measured by the instruments, CBI (circles) and ACBAR (squares). The
measurements have been converted from their values at the particular
instrument observing frequency to 300 GHz after subtraction of the primordial CMB contribution (see text for details) by assuming that they are
entirely due to the SZ effect and by then using the spectral SZ
frequency dependence (see equation \ref{equ:tszspec}). \label{fig:ps_lcdm}}
\end{center}
\end{figure}

\begin{figure}
\begin{center}
\includegraphics[scale=0.42]{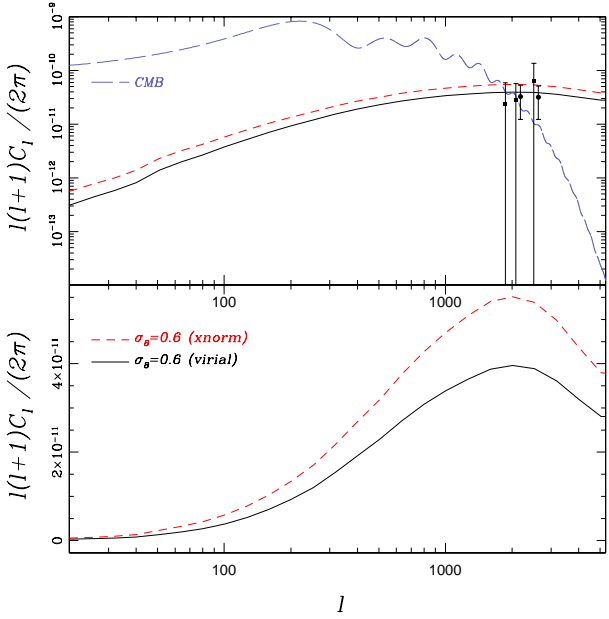}
\caption{The same as Figure \ref{fig:ps_lcdm} for the $\tau$CDM
simulations.
\label{fig:ps_tcdm}}
\end{center}
\end{figure}

Based on the modelling described above, full-sky maps of the SZ effects
are constructed. Patches of the Compton $y$ maps of the thermal SZ
effect for different values of $\sigma_8$ and different M-T relations
are shown in Figures \ref{fig:tszlcdmrel} and \ref{fig:tsztcdmrel} for
the $\Lambda$CDM and $\tau$CDM cosmology respectively. The models have
been chosen to cover a wide range of parameter space. In particular,
models which are supported by recent observations, such as the WMAP
mission and X-ray cluster and weak lensing measurements, have been
included. Moreover, the model choice has been
influenced by observational data of the CMB at high angular multipoles
($1500 < l < 3000$). To visualise the differences occurring from
changes in $\sigma_8$ and of the M-T relation, the same sky
patch is shown for the $\Lambda$CDM and $\tau$CDM models
respectively. Due to different structure formation histories the
surface density of extended low redshift foreground clusters is larger
in the $\tau$CDM model, whereas the $\Lambda$CDM models possess more
high redshift background clusters.

In Figures \ref{fig:ps_lcdm} and \ref{fig:ps_tcdm} the primordial CMB
normalised to WMAP measurements at low multipoles and thermal SZ power spectra (in the following referred to as SZ
 power spectra) are plotted at 300 GHz. The included points with error
 bars are observational results from ACBAR (Kuo et al. 2002) and CBI (Mason et al. 2001). The measurements
 have been converted from the particular instrument observing
 frequency to 300 GHz after subtraction of the primordial CMB
contribution by assuming that the remaining measured excess is entirely
 due to the SZ effect and by using the spectral SZ frequency
 dependence (see equation \ref{equ:tszspec}). The fraction contributed
by the primordial CMB has been estimated from an extrapolation of the
WMAP results best-fitting primordial CMB power spectrum to higher multipoles,
which has been normalised to the
height of the first acoustic peak. Further, it has been assumed that the
primordial CMB estimate is
free of error. Note that, at 1$\sigma$ level, no excess is required to
explain the ACBAR measurements. 
However, under the made
assumptions the measured values
of ACBAR and CBI are consistent with each other, the SZ interpretation
and the
primordial CMB contribution estimated on the basis of the WMAP measurements. The
consistency of the CBI and ACBAR excesses has also
been found by Goldstein et al.\,(2003) by simultaneously fitting
primordial CMB and SZ power spectra to the observations without using
WMAP constraints. While the
high-$\sigma_8$ $\Lambda$CDM model SZ power spectra fit the
observed excess measurements well, the lower $\sigma_8$ $\Lambda$CDM models tend to
 lie below the excess measured by CBI and ACBAR. This finding is
consistent with Bond et al.\,(2002) who obtained $\sigma_8 = 1.0$ using
SZ power spectra from hydrodynamical simulations to fit the CBI excess
for $\Lambda$CDM
cosmologies. The $\tau$CDM
 $\sigma_8 = 0.6$ SZ power spectra slightly overpredict the SZ power compared to
the observed excess (but they are excluded by less than 1$\sigma$). Nevertheless, for the presented work it is
 important to include a wide range of models in order to study model
 dependent selection effects of the employed cluster detection
 algorithm.

\subsection{The SZ power spectrum relation to $\sigma_8$}

\begin{figure}
\begin{center}
\includegraphics[scale=0.42]{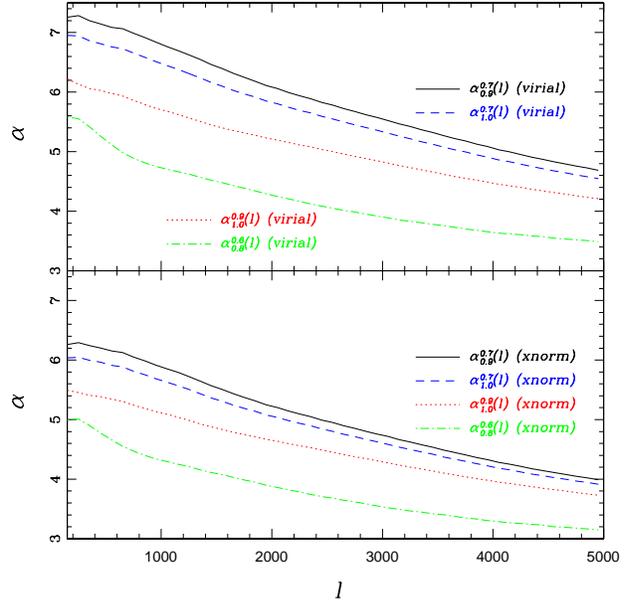}
\caption{The dependence of the power law index $\alpha$ on the
multipoles $l$ for the virial (upper panel) and the M-T relation
obtained from X-ray observations (lower panel) for our modelling of
the SZ sky. \label{fig:sig8_alpha}}
\end{center}
\end{figure}

Since the value of $\sigma_8$ has a significant effect on the
abundance of galaxy clusters and thus on the measured power of SZ anisotropies,
in the following the dependence of the thermal SZ power spectrum
amplitude on $\sigma_8$ is investigated on the basis of the performed
simulations. Predictions by various authors (Komatsu \& Kitayama 1999;
Seljak et al.\,2001; Zhang \& Pen 2001; Komatsu \& Seljak 2002; Sadeh
\& Rephaeli 2004) expect
$C_l \propto \left( \Omega_bh \right)^2 \sigma_8^{4-7}$. Based on such
predictions Bond et al.\,(2002) estimated $\sigma_8 \approx 1.0$ from
the excess power at high $l$ values measured by the CBI experiment
using hydrodynamical simulations of the SZ effect.

Most of the previous work on the SZ power spectrum dependence on
$\sigma_8$ utilised either a standard
Press-Schechter mass function (Press \& Schechter 1976) or
hydrodynamical simulations of small sky patches to obtain SZ power
spectra. Komatsu \& Seljak (2002) used the Jenkins mass function
(Jenkins et al.\, 2001), on which also the mass function of Evrard et
al.\,(2002) is built, and a cluster pressure profile fitted to
hydrodynamical simulations to evaluate the dependence of the SZ power
spectrum coefficients on cosmological parameters, especially
$\sigma_8$, for the Poissonian contribution at small angular scales. Being
based on the HV simulations our simulations ensure a realistic
modelling of the clustering contribution of galaxy clusters to the SZ
power spectrum.

Moreover, the mass cut can affect
the SZ power spectrum estimation. For the simulated $\Lambda$ and
$\tau$CDM models we performed a mass cut at $M_{cl}\approx 3\times
10^{13}\hMsol$. Komatsu \& Seljak (2002) find that a mass cut as low as
$5\times 10^{12}\hMsol$ is sufficient to obtain the SZ power spectrum
independently of mass boundaries for scales up to $l \approx
10^{4}$. For such low `cluster' masses the applicability of the
scaling relations as derived in section \ref{sec:szsim} is questionable, since
observations indicate a steepening of the M-T relation at mass scales
of the order of $1\times 10^{13}\hMsol$ (Finoguenov et al. 2001;
Borgani et al. 2002). On the
angular scales of interest the performed mass cut suffices to ensure
independence of the SZ power spectrum. This has been tested by a
comparison of the SZ power spectra obtained by performing different
mass cuts and on this basis the performed mass cut has been found to
be sufficient. For example, if we raise the mass cut from
$3\times 10^{13}\hMsol$ to $5\times 10^{13}\hMsol$, a decrease
of the SZ power spectrum amplitude at $l=5000$ of approximately 10 per
cent is found. Anyway, since most of the
clusters of the Planck survey will be unresolved (see section \ref{sec:rdist}),
Planck's ability to study substructures and cluster physics will be
(very) limited. Therefore, small mass clusters are mainly included to
provide a realistic SZ background noise.

Nevertheless, we present in the following an estimation based on our
cluster modelling of the dependence of the thermal SZ power spectrum
amplitude on $\sigma_8$ over a wide
range of angular scales $l$. Expressing
the relation between the SZ power spectrum amplitude and $\sigma_8$ by
\begin{equation}
 C_l \propto \sigma_8^{\alpha},
\end{equation}
$\alpha$ is obtained for two particular realisations of different
$\sigma_8$ by
\begin{equation}
\alpha(l) =
\frac{\mbox{log}\left(C_l(\sigma^{(1)}_8)/C_l(\sigma^{(2)}_8)\right)}{\mbox{log}\left(\sigma^{(1)}_8/\sigma^{(2)}_8\right)}.
\end{equation}
In Figure \ref{fig:sig8_alpha} the dependence of $\alpha$ on $l$ is
plotted for different combinations of $\sigma^{(1)}_8$ and
$\sigma^{(2)}_8$ for the used M-T relations (see section
\ref{sec:mtrel}).

Given our modelling of the SZ effect, $\alpha(l)$ is found to exhibit a
significant negative slope. While for the $\Lambda$CDM models at low
multipoles $\alpha$ adopts values of approximately 6 to 7, on small
angular scales ($l \geq 3000$) it has fallen to values of at least a
unit less. This behaviour can explain the wide range of values found in the literature
for the power-law index
$\alpha$. Sadeh \& Rephaeli (2004) obtained from analytical estimates of
the SZ power spectrum using the cluster gas profile of equation (\ref{equ:truncprojpro})
$C_l \propto \sigma_8^{4}$ at high $l$s, while previous studies at
lower $l$s estimated $C_l \propto \sigma_8^{6-7}$ (see e.g. Komatsu \&
Kitayama 1999). Thus it has to be taken into account for which
multipole range $\alpha$ is evaluated. Partially, this behaviour is
due to the clustering contribution which is most significant on large
angular scales and has been shown to possess a stronger scaling with
$\sigma_8$ (Komatsu \& Kitayama 1999). 

In the case of hydrodynamical
simulations the contribution due to substructure, which is expected to
increase the SZ power on small angular scales, possibly modifies the behaviour of the
power-law index $\alpha$ at high multipoles. As some
hydrodynamical simulations have been found to disagree with each other, it is at present impossible to
make robust predictions. For example, the
mesh-based GADGET code of Springel, White \& Hernquist (2001) seems to
predict more small-scale power than the simulations of da Silva et
al. (2001) using HYDRA, a SPH code. Since the SPH simulations of Bond
et al.\,(2002) agree in shape with Springel, White \& Hernquist (2001),
the disagreement is unlikely to be dependent on the differences of the applied
methods. The different shape of the high-$l$ SZ power spectra of
various hydrodynamical simulations could be due to a resolution
effect. The higher the resolution of the simulation the more
substructure is revealed which increases the amplitude of the power
spectrum at high multipoles and also flattens its peak. da Silva et
al.\,(2004) state that they have not yet attempted a resolution study
of their code. Anyway, since hydrodynamical simulations are computationally
expensive it is currently unfeasible to simulate volumes of the size needed for
realistic full-sky simulations of Planck observations.

Furthermore, in the case of the M-T relation derived from X-ray
measurements the power-law index systematically takes on smaller
values than for the virial M-T relation. This is due to the difference
of the cluster Compton $y$ profile induced by the change in
temperature corresponding to a fixed cluster mass. Since the cluster
radius and central electron gas density scale only with cluster mass
and redshift (see section \ref{sec:szsim}), the $y$ profile of the cluster is
altered by the increase of $T_e$. In our modelling the outskirts of the cluster are
more affected by this modification of the profile than its core and
therefore the effect is more significant at low multipoles. 

\begin{figure}
\begin{center}
\includegraphics[scale=0.42]{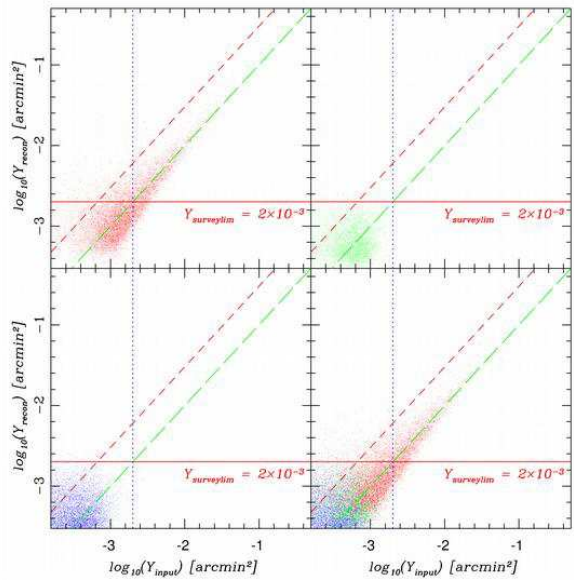}
\caption{Scatter plot of the reconstructed versus the actual cluster
Comptonization parameter $Y$ for clusters which have been matched to a
SZ peak detection by the described algorithm (see section \ref{sec:cldetec}) for the fiducial
$\Lambda$CDM model (see Table \ref{tab:param}). Based on the cluster detection
algorithm the upper left, the upper right and the lower left panel
show the scatter of the reconstructed cluster fluxes for clusters with
a peak detection of 5, 4 and 3 times $\sigma_{noise}$. The long dashed
green line gives the optimal reconstruction. The horizontal red line
represents a conservative survey $Y$ limit estimate. It is chosen to
lie well above the `confusion plateau' level. The dotted vertical blue
line shows the same flux limit for $Y_{input}$ ($Y_{input}\approx
2\times10^{-3}$arcmin$^{2}$). Due to cluster-cluster confusion even
low flux clusters of $Y\approx 1\times10^{-4}$arcmin$^{2}$ can have an
associated $Y_{recon}$ of several $1\times10^{-4}$arcmin$^{2}$. The
dashed red line gives a tolerant upper limit for when a
cluster-SZ peak match is taken to be correct. This
has been estimated from the skewness of the scatter around the optimal
cluster flux reconstruction and is chosen to be $3\times Y_{input}$
(see text for detailed explanation). The lower right panel shows all
detections.
\label{fig:yscat_lcdm}}
\end{center}
\end{figure}

As Figure \ref{fig:sig8_alpha} shows the power-law
index $\alpha$ takes on a lower value in the case of the $\tau$CDM
models. This agrees with the analytic expectations derived in
Komatsu \& Seljak (2002).
Moreover, the peak of the SZ power spectrum is shifted towards
smaller angular scales for smaller values of $\sigma_8$ (see lower
panels of Figures \ref{fig:ps_lcdm} and \ref{fig:ps_tcdm}) due to the rescaling of cluster
masses (see equation \ref{eq:dlnsig8dlnM}) and thus resizing of the clusters.

\subsection{The reconstructed SZ flux number count}
\label{sec:planckclnum}

First we consider the concordance cosmological model, the $\Lambda$CDM
model, which is supported by the WMAP measurements (Bennett et al.\,2003),
together with the virial M-T relation as our fiducial
model. In particular the adopted cosmological parameters are $\omm = 0.3$,
$\oml = 0.7$, $H_0 = 100 \, h \,$km\,s$^{-1}$Mpc$^{-1}$ with $h =
0.7$, $\Omega_b = 0.04$, $n = 1$ and $\sigma_8 = 0.9$ since these model
parameters yield a reasonable fit to the current strong cosmological
constraints and thus provide a robust basis for making realistic
predictions. Furthermore, these parameters agree with the $\Lambda$CDM
HV N-body simulations used to obtain the thermal and kinematic SZ
maps.

Figure \ref{fig:yscat_lcdm} shows the scatter of the
reconstructed SZ cluster integrated Comptonization parameters
versus their real ones. The reconstructed $Y$s have been
obtained by applying the algorithm as described in section
\ref{sec:cldetec}. Only
clusters whose real Comptonization parameter $Y$ lies above
$1\times 10^{-4}$arcmin$^{2}$ have been matched to detected
SZ peaks. This assumed lower sensitivity limit of $Y$ has been suggested by
analytical considerations (see e.g. Bartelmann 2001). In Figure
\ref{fig:yscat_lcdm} the scatter of $Y$ of detected clusters is shown for
different peak detection thresholds (5, 4 and $3\times
\sigma_{noise}$). Even though the majority of detections with a peak
above 4 or 3$\times \sigma_{noise}$ scatters symmetrically
around the line of optimal detection and therefore can be believed to
be matched up correctly, there is also a significant fraction which is
scattered upwards to larger fluxes. These detections are either false
matches or - which is more often the case - upscattered
due to cluster-cluster confusion. Thus for lower real cluster fluxes
the reconstructed integrated Comptonization parameters level off. The shown survey limit estimate
($Y_{surveylim} = 2\times 10^{-3}$arcmin$^{2}$) is chosen to lie well above this
flux `plateau' to ensure that the reconstructed flux distribution
represents closely the real one. As Figure \ref{fig:yscat_lcdm}
illustrates, the contribution of $3-$ and $4-\sigma_{noise}$ peak
detections above the sensitivity estimate for Planck is
small. 
 
\begin{figure*}
\begin{center}
\subfigure{\includegraphics[angle=0,width=0.45\textwidth]{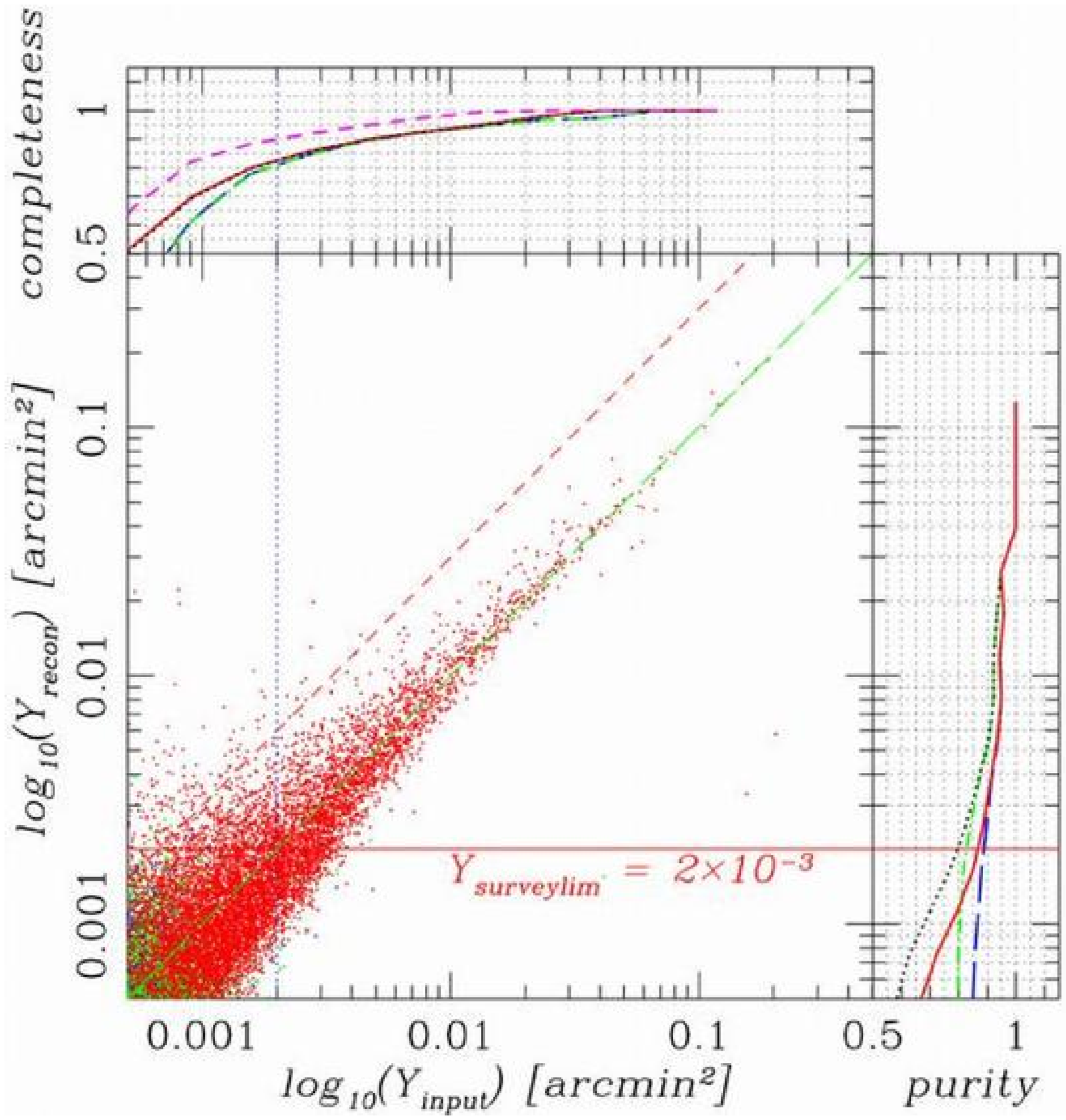}}
\subfigure{\qquad\includegraphics[angle=0,width=0.45\textwidth]{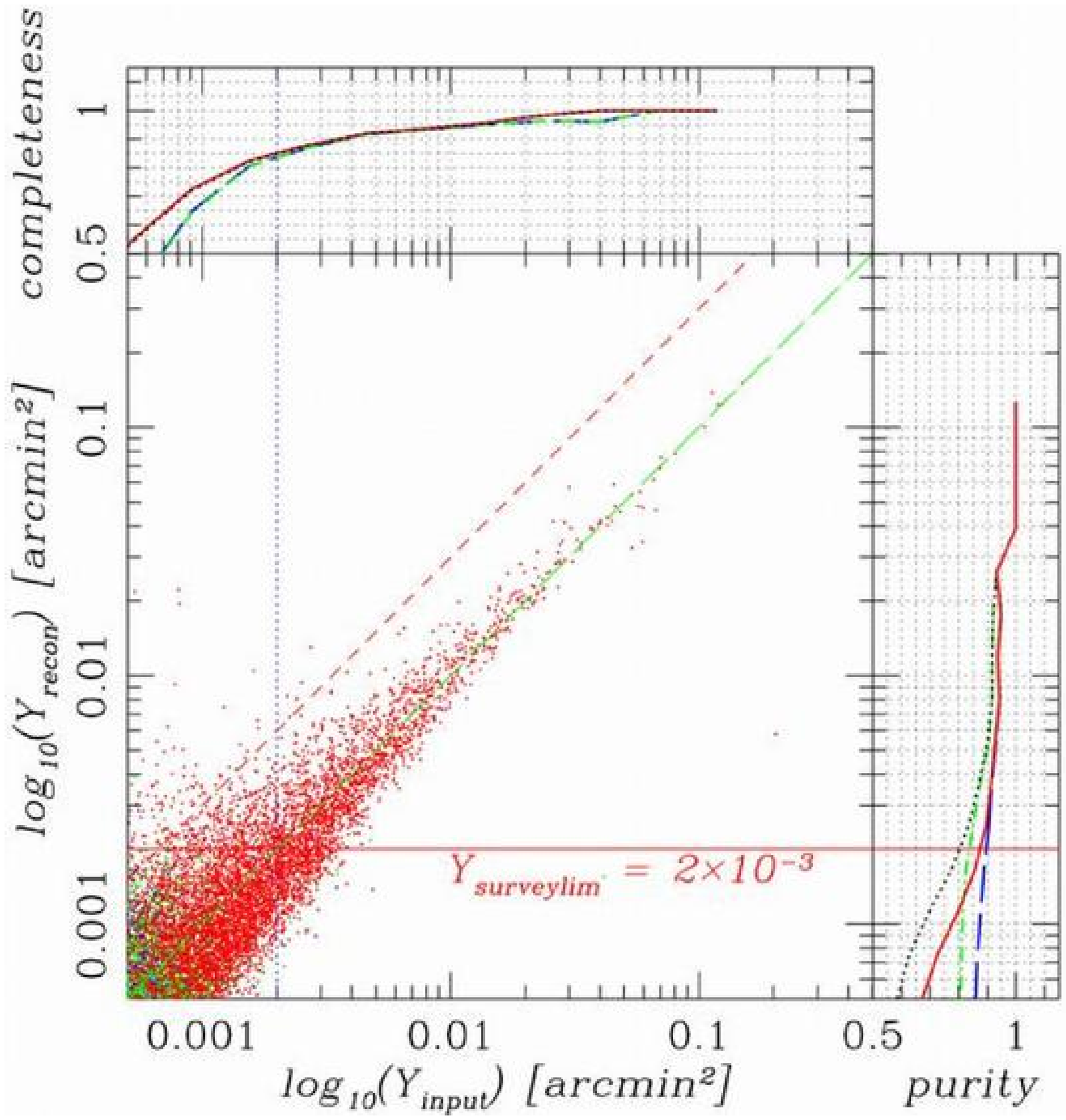}}
\caption{In the left panel the same scatter as in Figure
\ref{fig:yscat_lcdm} is shown (for comparison), while a Galactic cut of $\pm
20^{\circ}$ Galactic latitude has been performed to obtain the scatter
in the right panel. In this Figure the completeness and purity of the
reconstructed cluster sample are shown. In the purity and
completeness plots the red solid line represents the case in
which all detections with a peak amplitude above $3\times
\sigma_{noise}$ are taken into account, the blue long-dashed the case
which utilises only detections whose peak amplitude is $\ge 5\times
\sigma_{noise}$. The black dotted line and the green short-dashed line
show both cases when the matches which scatter above the red dashed
diagonal line are not approved as detections. In the left panel the magenta short-dashed line gives the completeness
of the survey when multiple clusters are allowed to
be matched to a single detection peak. \label{fig:yscat_gc_lcdm}}
\end{center}
\end{figure*}

\begin{table*}
\begin{center}
\begin{tabular}{||l||r||r|r|r|r|r||} \hline
\emph{Model} & \multicolumn{6}{c||}{\emph{$\Lambda$CDM}}\\ \hline
\emph{$\sigma_8$ and M-T relation} & 0.7 (virial) & 0.7 (xnorm) & 0.9
(virial) & 0.9 (xnorm) & 1.0 (virial) & 1.0 (xnorm)\\ \hline
\emph{$n_{cl}(Y_{input}\geq 5\times 10^{-3})$} & 219 & 405 & 959/642 &
1462 & 1575 & 2299\\ \emph{$n_{cl}(Y_{recon}\geq 5\times 10^{-3})$} &
260 & 497 & 957/662 & 1493 & 1528 & 2266\\ \hline
\emph{$n_{cl}(Y_{input}\geq 4\times 10^{-3})$} & 307 & 573 & 1348/903
& 2053 & 2206 & 3206\\ \emph{$n_{cl}(Y_{recon}\geq 4\times 10^{-3})$}
& 379 & 699 & 1325/905 & 2062 & 2093 & 3186\\ \hline
\emph{$n_{cl}(Y_{input}\geq 3\times 10^{-3})$} & 480 & 899 & 2083/1403
& 3207 & 3350 & 4801\\ \emph{$n_{cl}(Y_{recon}\geq 3\times 10^{-3})$}
& 625 & 1128 & 2000/1371 & 3193 & 3180 & 4920\\ \hline
\emph{$n_{cl}(Y_{input}\geq 2\times 10^{-3})$} & 930 & 1741 &
3781/2561 & 5770 & 5941 & 8701\\ \emph{$n_{cl}(Y_{recon}\geq 2\times
10^{-3})$} & 1478 & 2360 & 3659/2532 & 5998 & 5784 & 9035\\ \hline
\emph{$n_{cl}(Y_{input}\geq 1\times 10^{-3})$} & 2791 & 5263 &
10228/6735 & 16114 & 15773 & 23826\\ \emph{$n_{cl}(Y_{recon}\geq
1\times 10^{-3})$} & 5506 & 7764 & 10744/7405 & 16626 & 15736 &
23358\\ \hline
\end{tabular}
\caption{Cluster number counts above certain limits of the
Comptonization parameter $Y$ for the $\Lambda$CDM models with
different $\sigma_8$ and for the two utilised M-T relation normalisations. In
each case the expected input and recovered number of clusters above
the assumed $Y$ limit (in arcmin$^{2}$) is given. In the virial
$\sigma_8 = 0.9$ case the stated numbers are for a full-sky survey as
well as for one with a Galactic cut of $\pm
20^{\circ}$ Galactic latitude (utilising 2/3 of the sky). \label{tab:cl_flux_dist_lcdm}}
\end{center}
\end{table*}

\begin{table}
\begin{center}
\begin{tabular}{||l||r||r||} \hline
\emph{Model} & \multicolumn{2}{c||}{\emph{$\tau$CDM}}\\ \hline
\emph{$\sigma_8$ and M-T relation} & 0.6 (virial) & 0.6 (xnorm)\\ \hline \emph{$n_{cl}(Y_{input}\geq 5\times
10^{-3})$} & 3043 & 4620\\ \emph{$n_{cl}(Y_{recon}\geq
5\times 10^{-3})$} & 3070 & 4800\\ \hline
\emph{$n_{cl}(Y_{input}\geq 4\times 10^{-3})$} & 4134 & 6243\\ 
\emph{$n_{cl}(Y_{recon}\geq 4\times 10^{-3})$} & 4194 & 6602\\ \hline \emph{$n_{cl}(Y_{input}\geq 3\times 10^{-3})$}
& 6030 & 9264\\ \emph{$n_{cl}(Y_{recon}\geq 3\times
10^{-3})$} & 6200 & 9803\\ \hline
\emph{$n_{cl}(Y_{input}\geq 2\times 10^{-3})$} & 10441 & 16051\\ \emph{$n_{cl}(Y_{recon}\geq 2\times 10^{-3})$} & 10523 &
16729\\ \hline \emph{$n_{cl}(Y_{input}\geq 1\times
10^{-3})$} & 25410 & 39689\\
\emph{$n_{cl}(Y_{recon}\geq 1\times 10^{-3})$} & 24160 & 36155\\ \hline
\end{tabular}
\caption{The same as Table \ref{tab:cl_flux_dist_lcdm} for the
$\tau$CDM model. \label{tab:cl_flux_dist_tcdm}}
\end{center}
\end{table}

The reconstructed fluxes of some clusters are upscattered due to
cluster-cluster confusion and even for the $5-\sigma_{noise}$
detections the scatter around the optimal
reconstruction is slightly skewed. Hence, there happen to be some
outliers whose reconstructed $Y$ is several times their $Y_{input}$. One way to deal with this is the
introduction of a liberal upper $Y$-limit that ensures that these
questionable cluster reconstructions can be excluded from the sample.
Thus by enforcement of this limit clusters with a reconstructed
$Y_{recon}$ above $3\times Y_{input}$ are regarded as false
detections. On first sight this choice seems to be quite high, but one
has to keep in mind that clusters confusing each other often have
comparable masses and are situated next to each other in redshift
space. In the following estimates of detection purity and
completeness of the presented
cluster detection method are given with and without taking account of
this upper limit. Furthermore, these characteristics of the
algorithm are shown for all cluster detections with associated peaks above
$3\times\sigma_{noise}$ as well as only for the $5-\sigma_{noise}$
peak detections. Since the reconstructed integrated Comptonization
parameters of high-flux clusters scatter almost symmetrically around the line
of optimal flux reconstruction, this indicates that the estimate of
the total cluster flux obtained by our algorithm for clusters above
$Y_{surveylim} = 2\times 10^{-3}$arcmin$^{2}$ is fairly unbiased.
The levelling off of the reconstructed $Y$s at lower fluxes is
mainly due to clusters being confused with each other.

The completeness and purity obtained by the applied algorithm are plotted in
Figure \ref{fig:yscat_gc_lcdm} for the full-sky simulation and for a
Galactic cut of Galactic latitude $|b|\leq 20^{\circ}$. The Galactic
cut has been performed in order to evaluate the influence of Galactic
residuals in the thermal SZ reconstruction on the quality of the
obtained cluster number count. For the full-sky cluster number
count, the cluster survey is more than 83 per cent complete and the
purity exceeds 88 per cent above the estimated survey limit,
$Y_{surveylim} = 2\times 10^{-3}$arcmin$^{2}$, for all clusters without
enforcement of an upper matching limit. While the completeness is
unaffected by the introduction of the
proposed upper $Y$-limit, the purity estimate is reduced by
approximately 5 per cent. This is the case when all detected clusters
are considered as well as for only considering $5-\sigma_{noise}$ peak
detections.  In any case neither the purity nor the completeness fall
below 80 per cent for $Y_{surveylim} = 2\times 10^{-3}$arcmin$^{2}$ for a cluster
survey performed over the full sky.  In the case of the Galactic cut,
the completeness is increased by a few per cent, whereas the purity is
barely improved for clusters with $Y$ above $2\times 10^{-3}$arcmin$^{2}$. Thus
the Galaxy does not introduce significantly false detections into the
recovered cluster sample. Some clusters are missed due to Galactic
contamination. However, the improvement of the completeness is mostly
cancelled by the reduction in the number of detected clusters which
leads to an increase in the Poissonian sampling error of about the same
percentage at $Y_{surveylim} = 2\times 10^{-3}$arcmin$^{2}$.

Note that the resulting completeness and purity also depend on the
reliability of the cluster-SZ peak matching. Since the number of
clusters and SZ peaks which have to be matched is large, the matching
has to be automated. Even though an optimisation has been performed,
it occasionally happens that the algorithm fails to match clusters and
SZ peak detections correctly. The number of false detections and undetected
clusters are both increased if a SZ peak which actually corresponds to
a cluster is not matched up. This occurs, for example, in the
case in which two clusters are confused with each other by
beam-convolution and the SZ peak corresponding to both lies outside
the matching acceptance region. Therefore, the quoted values have to
be regarded as conservative lower limits of the cluster
finding algorithm. This is supported by the comparison of the number
of detected SZ peaks whose flux estimation exceeds the sensitivity limit and
the number of clusters above $Y_{surveylim} = 2\times 10^{-3}$arcmin$^{2}$ which
are contained in the map (see Tables \ref{tab:cl_flux_dist_lcdm} and
\ref{tab:cl_flux_dist_tcdm}).

\begin{figure}
\begin{center}
\includegraphics[scale=0.42]{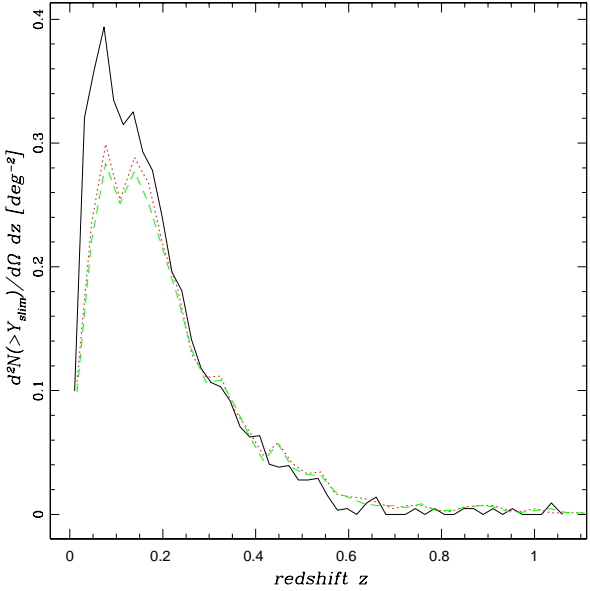}
\caption{The differential cluster number redshift distribution per
square degree for the HV cluster catalogue (black solid
line), for the full-sky recovered galaxy cluster survey (green dashed line) and
when performing a Galactic cut of $\pm
20^{\circ}$ Galactic latitude (red dotted line) for an assumed survey sensitivity limit of $Y_{surveylim}=
2\times 10^{-3}$arcmin$^{2}$. 
\label{fig:nzlim}}
\end{center}
\end{figure}

Figure \ref{fig:nzlim} shows the differential number redshift
distribution of clusters with and without a performed Galactic cut as compared to the expected one for $Y \geq 2\times
10^{-3}$arcmin$^{2}$. The largest difference between the expected distribution and
the recovered ones occurs at low redshifts. This is mainly due to
cluster-cluster confusion by beam-convolution. Due to line-of-sight
projection, an extended foreground cluster is more likely to overlap
with background clusters. Part of this
observed effect, however, is caused by
the reconstruction noise, since extended foreground clusters have
smaller signal-to-noise peak detections than high-redshift clusters
which appear more compact when projected along the line-of-sight. At
higher redshifts the differential reconstructed number counts
sometimes exceed those expected. This is due to clusters whose real fluxes are
just below the survey threshold and being upscattered due to confusion
of their SZ signal by projected overlapping clusters. The Galactic cut
distribution yields slightly better results than the full-sky one.

\begin{figure*}
\begin{center}
\subfigure[$\Lambda$CDM: $\sigma_8=0.7$
(virial)]{\label{fig:nonres:b}\includegraphics[angle=0,width=0.3\textwidth]{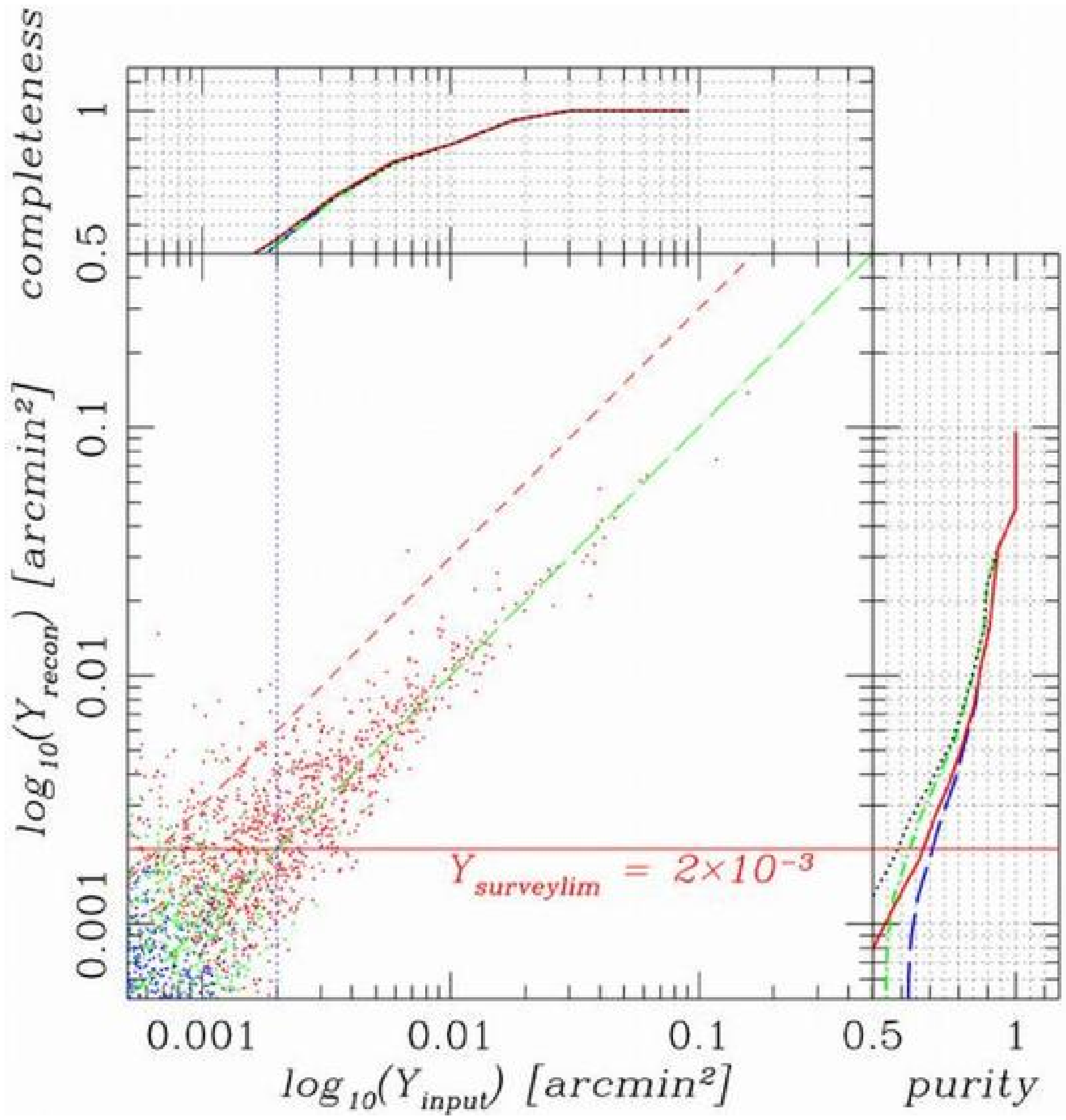}}
\subfigure[$\Lambda$CDM: $\sigma_8=1.0$
(virial)]{\label{fig:nonres:c}\includegraphics[angle=0,width=0.3\textwidth]{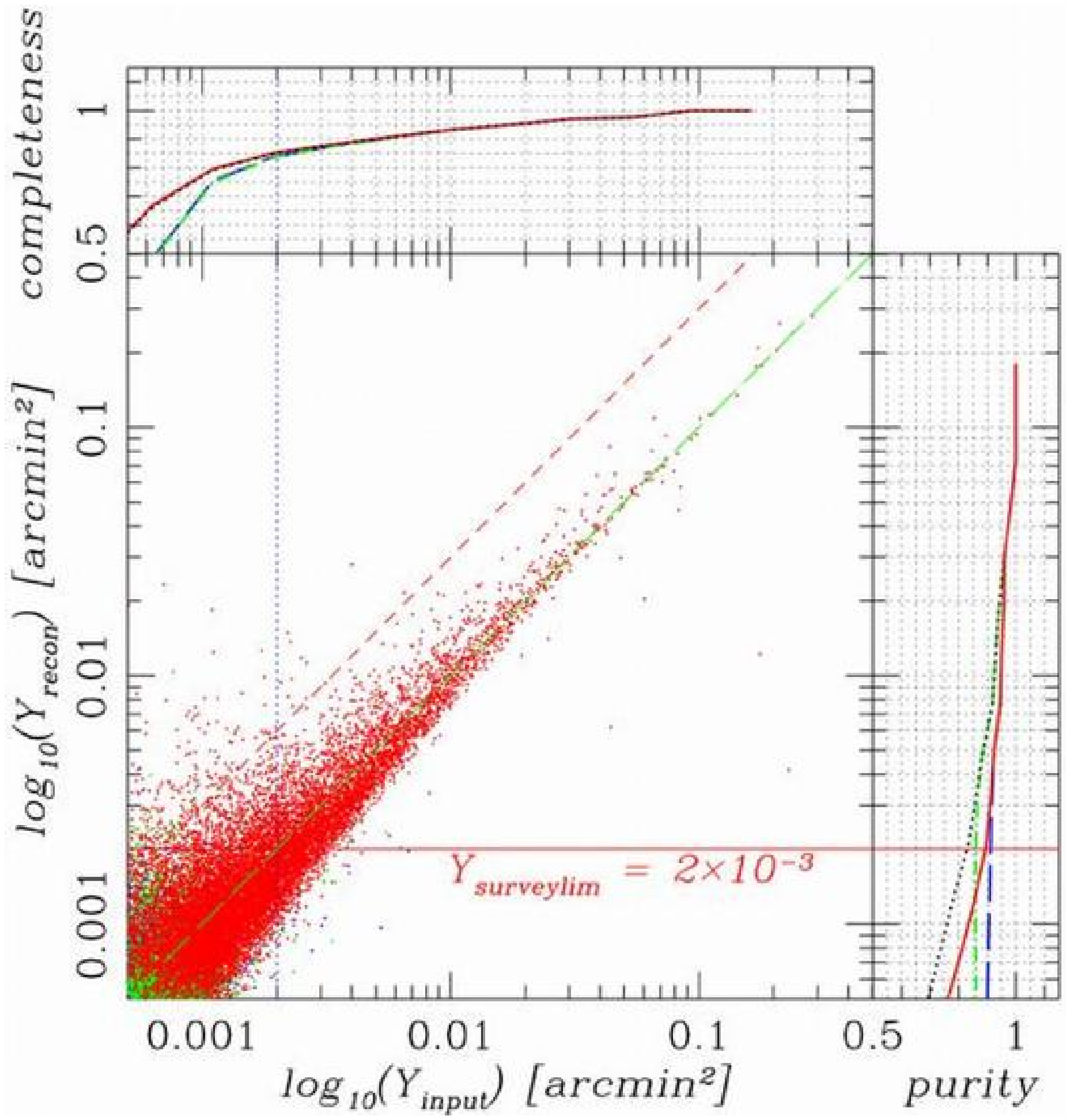}}
\subfigure[$\tau$CDM: $\sigma_8=0.6$
(virial)]{\label{fig:nonres:d}\includegraphics[angle=0,width=0.3\textwidth]{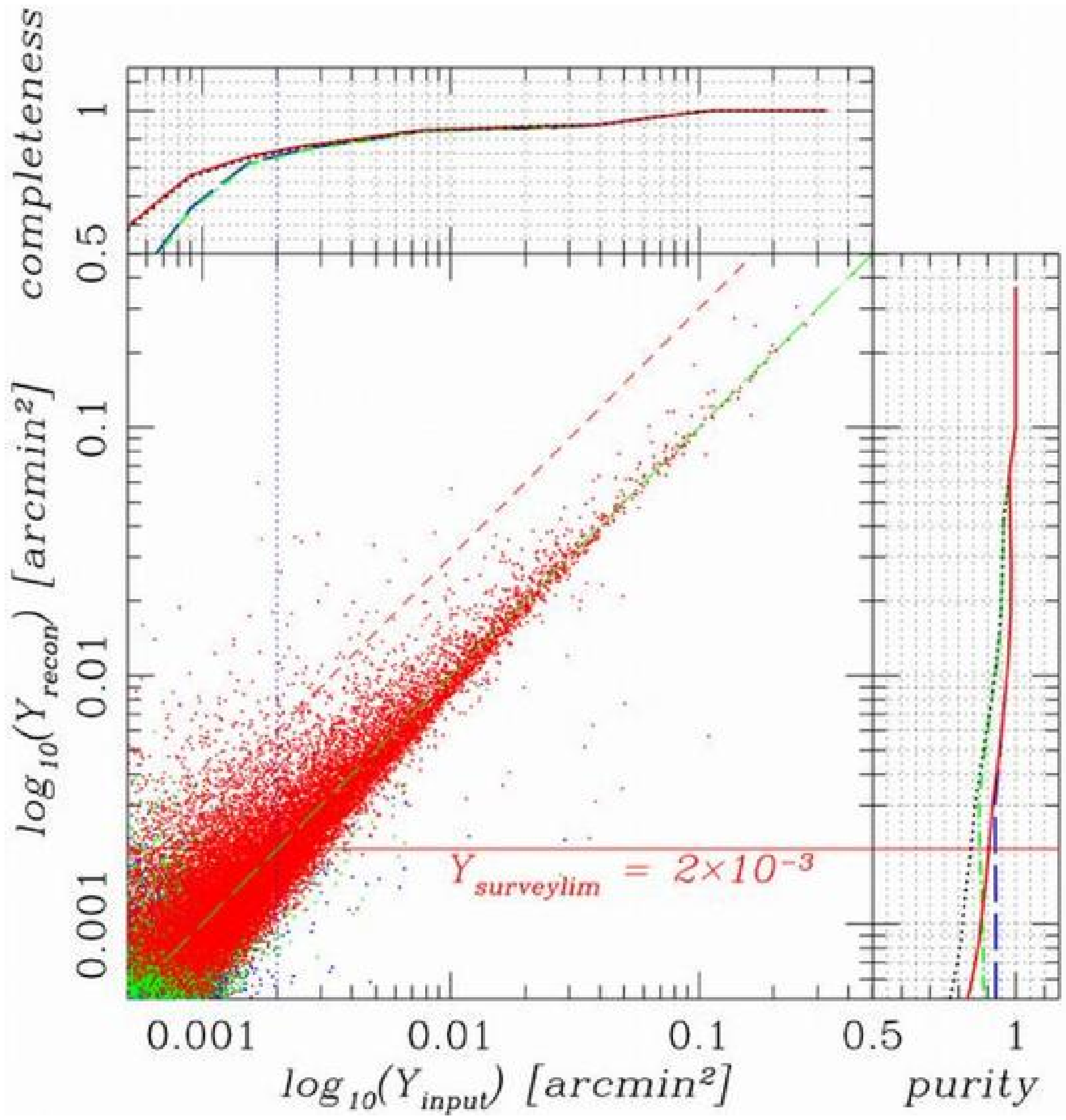}}
\subfigure[$\Lambda$CDM: $\sigma_8=0.7$
(xnorm)]{\label{fig:nonres:e}\includegraphics[angle=0,width=0.3\textwidth]{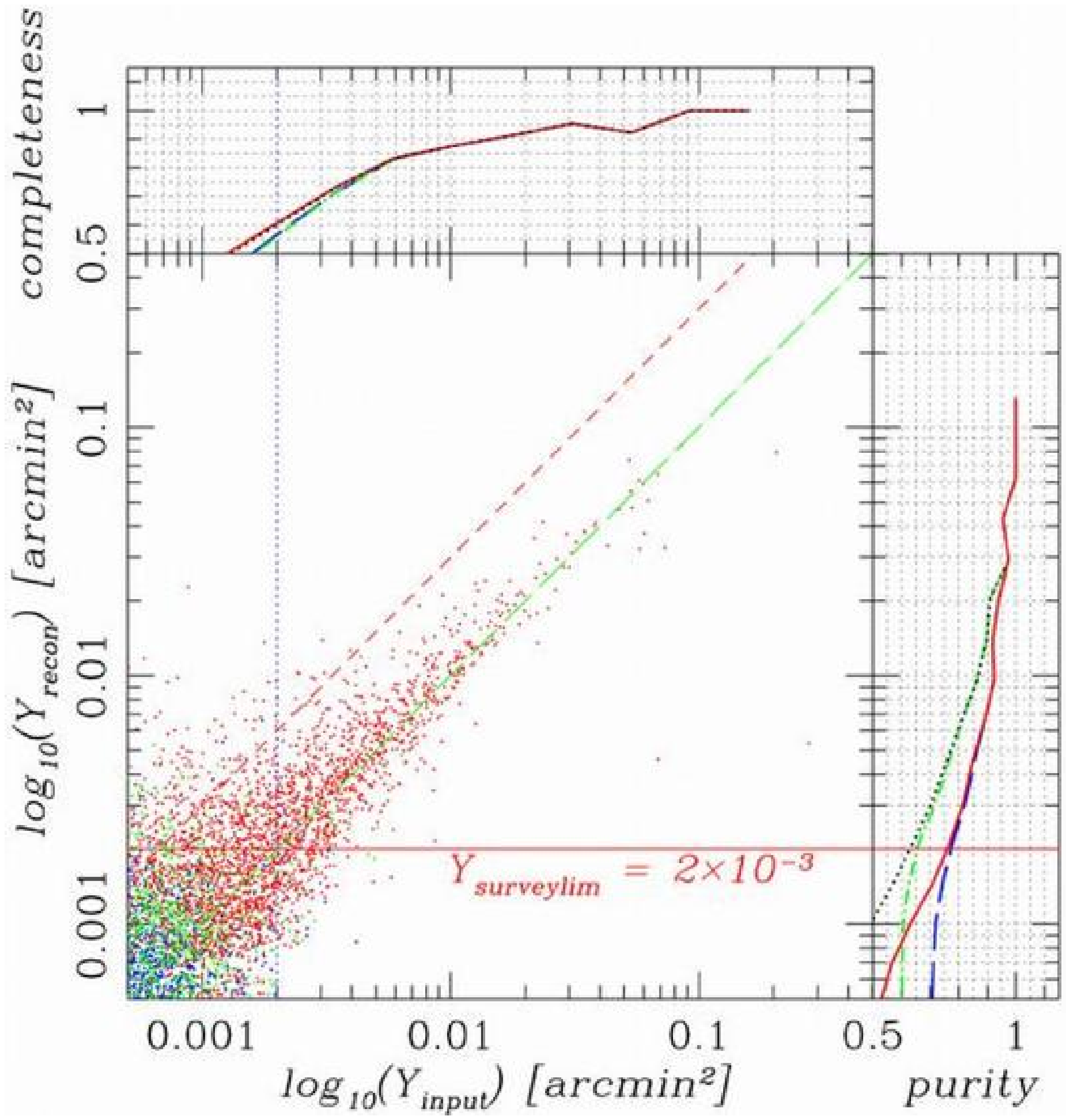}}
\subfigure[$\Lambda$CDM: $\sigma_8=1.0$
(xnorm)]{\label{fig:nonres:e}\includegraphics[angle=0,width=0.3\textwidth]{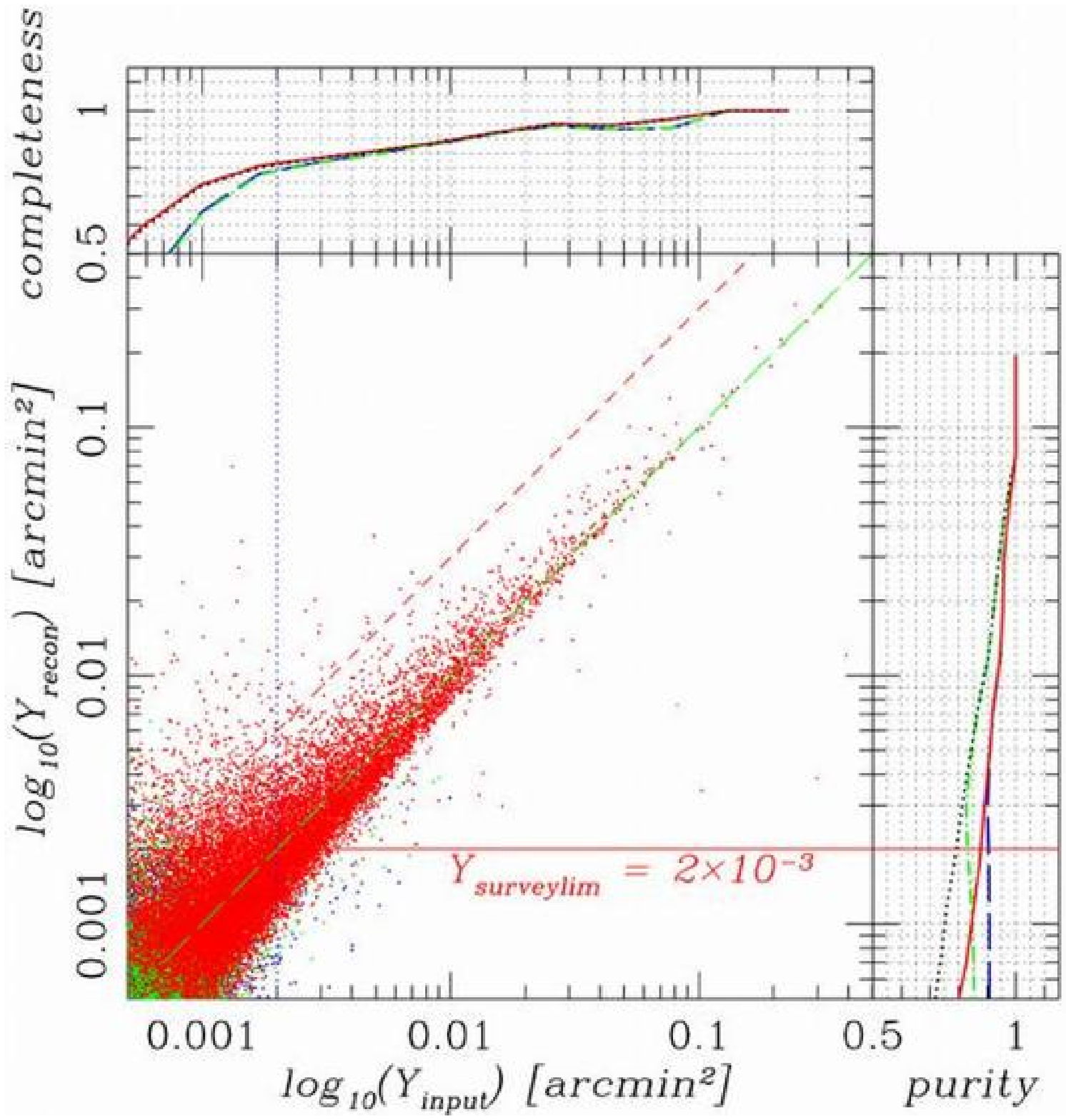}}
\subfigure[$\tau$CDM: $\sigma_8=0.6$
(xnorm)]{\label{fig:nonres:e}\includegraphics[angle=0,width=0.3\textwidth]{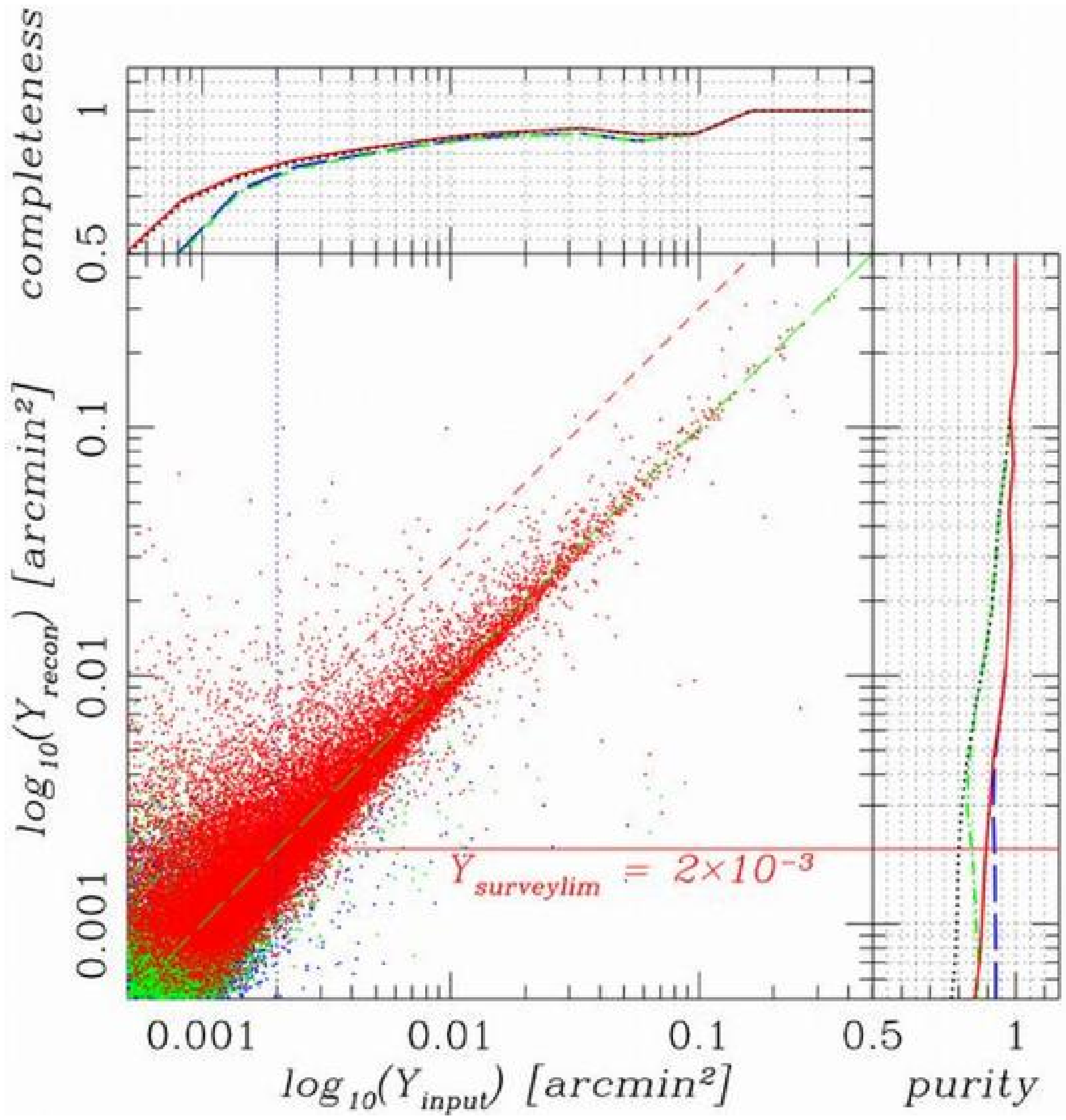}}
\subfigure[$\Lambda$CDM: $\sigma_8=0.9$
(xnorm)]{\label{fig:nonres:a}\includegraphics[angle=0,width=0.3\textwidth]{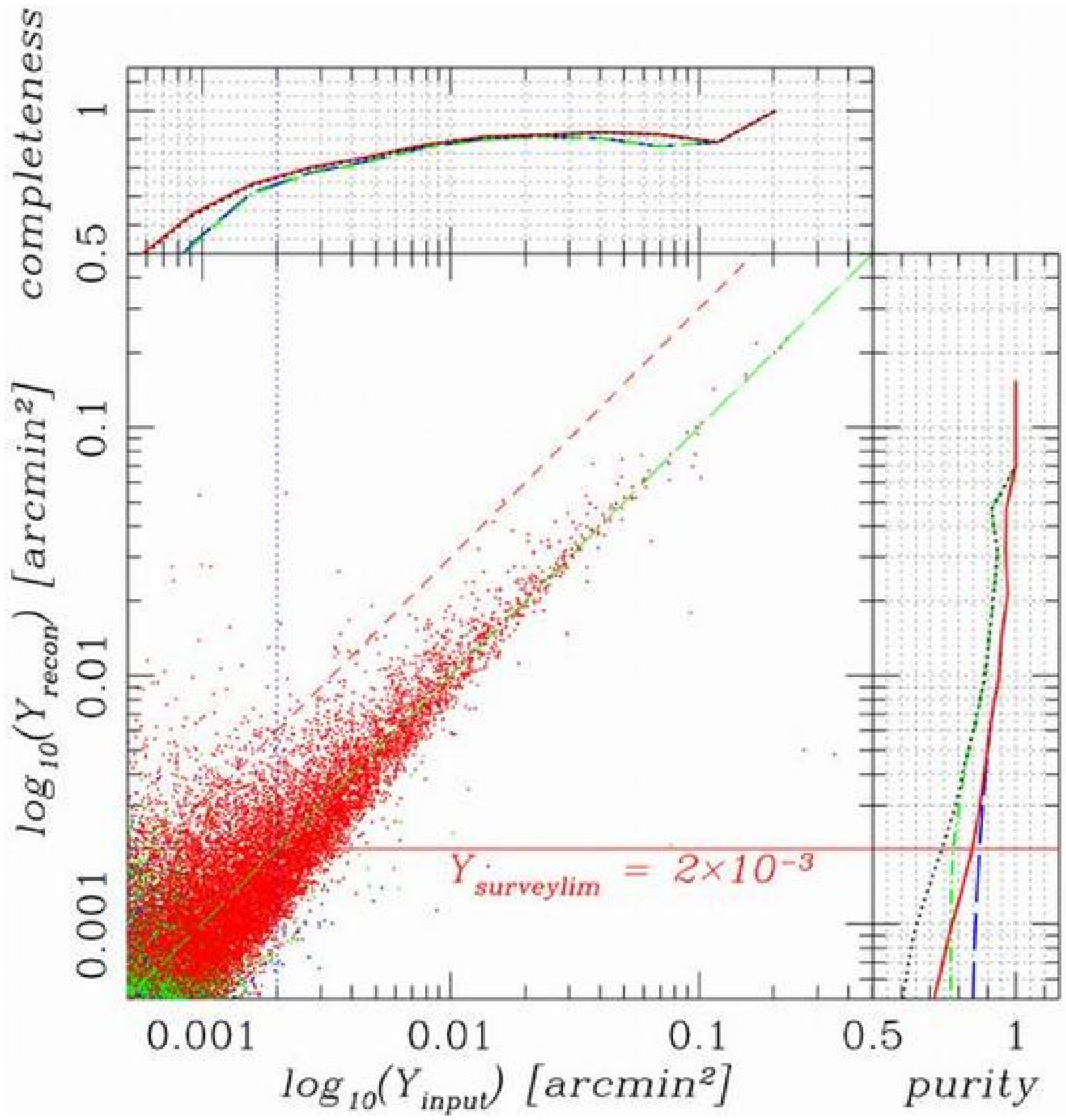}}
\caption{The scatter, completeness and purity for all performed
cosmological SZ realisations apart from the fiducial $\Lambda$CDM one are shown. \label{fig:yscat_all}}
\end{center}
\end{figure*}

In Figure \ref{fig:yscat_all} results of the application of the
algorithm to models with cosmological parameters which differ from the
concordance model are shown. This has been done to test the algorithm
which has been optimised for the fiducial $\Lambda$CDM cosmology. The
models are chosen to incorporate a wide range of $\sigma_8$ which is
still allowed by observations to ensure that in any case the method
delivers reliable results. The WMAP measurements on their own do not
put strong constrains on $\sigma_8$. Hence, $\sigma_8 = 0.7$ has been
taken as the lower limit of the allowed range as this value is
currently supported by X-ray cluster survey analyses (Allen et
al.\,2003). On the other hand weak lensing observations tend
to favour higher values (see e.g. Hoekstra et al.\,2002; Bacon et al.\,2003). Thus $\sigma_8 = 1.0$ may be considered as
an upper limit of the allowed range. Besides the favoured $\Lambda$CDM
model we include a flat model dominated by
non-relativistic matter without a cosmological constant ($\tau$CDM model) in our investigations.

We find, apart from the low-$\sigma_8$ $\Lambda$CDM model, the estimated
completeness as well as purity obtained for the additional models
are of equivalent or even improved quality in comparison to estimates
for the concordance model and the virial M-T relation.

In the case of the low-$\sigma_8$ $\Lambda$CDM models noise introduced
by the HSMEM itself and residual noise of other components become more
significant due to the lower SZ signal amplitude. The
completeness and purity for all cluster counts with $Y$ above the
survey limit are reduced to 56 and 68 per cent respectively (without
enforcement of an upper matching limit). Thus, the applied
algorithm does not quite suffice anymore to provide a reconstruction quality
from which tight constraints on cosmological parameters, $\sigma_8$
and $\Omega_m$, can be derived or which can establish a robust basis for
successful follow-up observations.

Moreover, for models with a larger variance of matter fluctuations on scales of
$8$ $h^{-1}$Mpc, the cluster confusion is more significant. Due to the mass
rescaling obtained by employing equation (\ref{eq:dlnsig8dlnM}) a cluster gets
associated with a larger mass and therefore has a larger extent. This
results in a more significant
cluster overlap when projecting clusters SZ signal onto the
two-dimensional full-sky map. As a result the completeness slightly
decreases due to increasing cluster-cluster
confusion. On the other hand the purity is slightly increased due to
the more dominant SZ signal. A similar effect is generally found when the
cluster temperatures are increased by a change of the M-T scaling
relation normalisation. Then due to beam-convolution neighbouring
clusters affect each other more strongly. Even though it can be argued that
in any case of cluster-cluster confusion a confirmation procedure would be able to match the contributing
clusters and even disentangle their contributions, these more optimistic considerations
have not entered the presented results.

Tables \ref{tab:cl_flux_dist_lcdm} and \ref{tab:cl_flux_dist_tcdm}
give the detected cluster numbers above a declining threshold for each
simulation. For comparison, the number of clusters whose flux
exceeds the particular threshold and has been obtained directly from
the cluster catalogues is also given. For $Y_{surveylim} = 2\times 10^{-3}$arcmin$^{2}$
we expect Planck to detect several 1000 clusters. If $Y_{surveylim}$
can be lowered to $1\times 10^{-3}$arcmin$^{2}$, Planck will find a total number
of clusters of the order $10^{4}$. Since the recovered and input
cluster number show good agreement even for $Y_{lim}=1\times 10^{-3}$arcmin$^{2}$,
it is likely that, when the matching of SZ peak detections with clusters is
even more optimised, a flux limit of $1\times 10^{-3}$arcmin$^{2}$ may be
reached.

So far we have taken into account the quality of the cluster flux
recovery to estimate a sensitivity limit for a Planck cluster
survey. Previous work often
considered a cluster to be detected when it could be successfully
matched with a peak in the reconstructed SZ map and allowed even
several clusters to be matched with a single SZ peak. It can be argued
that follow-up observations at higher resolution or at other
wavelengths are able to disentangle the contributions of several
clusters to an extended SZ feature and to confirm Planck cluster
detections. Such follow-up observations can be
performed, for example, by ground-based SZ telescopes, such as the interferometers
AMI (Kneissl et al.\,2001), Amiba (Lo et el.\,2000) and SZA (Mohr et
al.\,2002) in the northern hemisphere and the upcoming instruments in
the south, such as SPT\footnote{http://astro.uchicago.edu/spt} and ACT\footnote{http://www.hep.upenn.edu/$\sim$angelica/act/act.html}. If we assume that the contributions of
individual clusters can be separated, the completeness
is significantly increased (see Figure \ref{fig:yscat_lcdm}). The obtained cluster sample is 83 per cent complete
at $Y_{lim}=1\times 10^{-3}$arcmin$^{2}$ and still has a purity of 78
per cent.

\begin{figure}
\begin{center}
\includegraphics[scale=0.42]{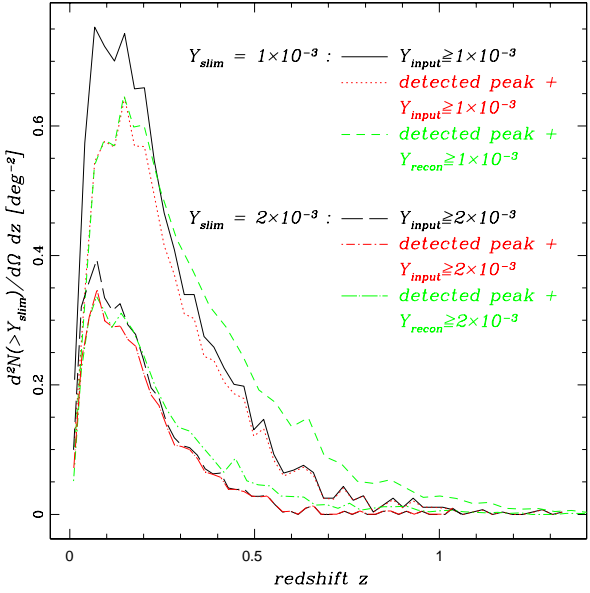}
\caption{The differential cluster number redshift distribution per
square degree for the HV input cluster catalogue and for the full-sky galaxy
cluster detections under the assumption that a cluster is detected
when its matching region includes a reconstructed SZ peak. Thus,
multiple clusters can be matched to a single peak. In the case of
the peak-detected clusters above a given flux survey limit the
distributions are shown for the true cluster flux as well as for the
flux associated with the peak matched to the cluster (different
distributions are indicated by different colour and line-styles). The
plot shows the
distributions for assumed survey sensitivity limits of $2\times 10^{-3}$arcmin$^{2}$ and $1\times 10^{-3}$arcmin$^{2}$. \label{fig:nzuplim}}
\end{center}
\end{figure}

For the fiducial $\Lambda$CDM model, Figure \ref{fig:nzuplim} compares
the effect of these assumptions on the cluster sample obtained
for the two considered sensitivity limits by the differential redshift
number distributions. It is found that due to cluster-cluster
confusion, especially at higher redshifts ($z \ge 0.3$), at which the
angular separation of clusters situated next to each other in redshift
space is reduced, the recovered distributions using the recovered
cluster fluxes exceed the expected cluster numbers significantly for
both survey limits. This is caused by an increased likelihood of
clusters with fluxes below the particular limit to be included in the
sample, since all clusters contributing to the connected peaked SZ
feature are taken into account. On the basis of Planck observations
analysed with the presented cluster detection method, it is not
possible to separate clusters contributing to a single SZ peak
detection. Thus, follow-up observations are unavoidable in order to
disentangle clusters and remove low-flux clusters contributing to a SZ
peak detection above the flux limit from the sample. Then it is
feasible to reach the distributions which are
given for the
cluster peak detections assuming knowledge of the true cluster
fluxes (see Figure \ref{fig:nzuplim}). Further, Figure
\ref{fig:nzuplim} shows that the recovered distributions (using
cluster input as well as recovered fluxes) of the higher survey flux
limit ($Y = 2\times 10^{-3}$arcmin$^{2}$) resemble the true input
distribution more closely than it is the case for the lower flux
limit, $Y_{lim}=1\times 10^{-3}$arcmin$^{2}$. In the following
the more conservative estimate of the sensitivity
limit, $Y_{surveylim} = 2\times 10^{-3}$arcmin$^{2}$, is assumed and
only single cluster-peak matches are allowed.

Based on the reconstructed number counts, the purity and completeness
estimate, we can examine how likely models exclude each other given
the count of a certain model and comparing it to the expected count of the
ones that yield similar cluster numbers above the conservative survey limit. This is first done by
assuming that our estimates of the percentage of impure and missing
clusters are standard deviations of a Gaussian
distribution centred on the reconstructed and expected cluster number respectively. Furthermore, we account for the cosmic variance in the
number of clusters. Assuming a M-T relation, we find that given
the recovered count of a particular model, each of the other models considered here can be excluded by at least $1.7\sigma$. This minimal
exclusion is obtained in the case of the reconstructed $\sigma_8=0.9$ $\Lambda$CDM
count when compared with the expected count of the $\sigma_8=1.0$
$\Lambda$CDM model. However, the $\sigma_8=0.6$
$\tau$CDM cluster count excludes the $\sigma_8=1.0$ $\Lambda$CDM model by $2.6\sigma$.
When we take our estimates of impurity and incompleteness to be
absolute values, which have an error themselves, then the probability
of exclusion of
the respective models with similar expected number counts exceed
$5\sigma$ in all cases. Since these estimates have been obtained by a large
number of single cluster detections and we can regard them as the
probability that a cluster above the survey limit is detected or a
real detection respectively, this assumption should be valid
and we can assume a Poissonian sampling error in the completeness and
purity. Nevertheless, we first considered the more conservative case
of the estimated incompleteness and impurity to be the first moments of a
Gaussian distribution of the expected and recovered number count.

As expected the cluster number exceeding a given flux limit is
increased for the M-T scaling relation normalised by X-ray
observations (xnorm). An uncertainty in the normalisation of the M-T
relation introduces extra degeneracies besides the ones present between
cosmological parameters. For example, the
cluster number yield of the $\Lambda$CDM $\sigma_8=0.9$ model
employing the xnorm M-T scaling relation is comparable to the
one obtained for the $\Lambda$CDM $\sigma_8=1.0$ cosmology using the
virial M-T relation.

\subsection{Cluster size distribution and cluster resolution}
\label{sec:rdist}

To study galaxy cluster physics resolution is important. For the
Planck satellite, the large
beam sizes mean that most clusters will be unresolved. Nevertheless, a minor fraction of the cluster sample,
nearby massive clusters, will be resolved. Besides giving an estimate of the
expected percentage of resolved clusters contained in the sample above the survey
limit estimate ($Y_{surveylim} = 2\times 10^{-3}$arcmin$^{2}$), we further
investigate how well cluster radii are recovered by the applied
cluster finding algorithm. 

\begin{figure}
\begin{center}
\includegraphics[scale=0.42]{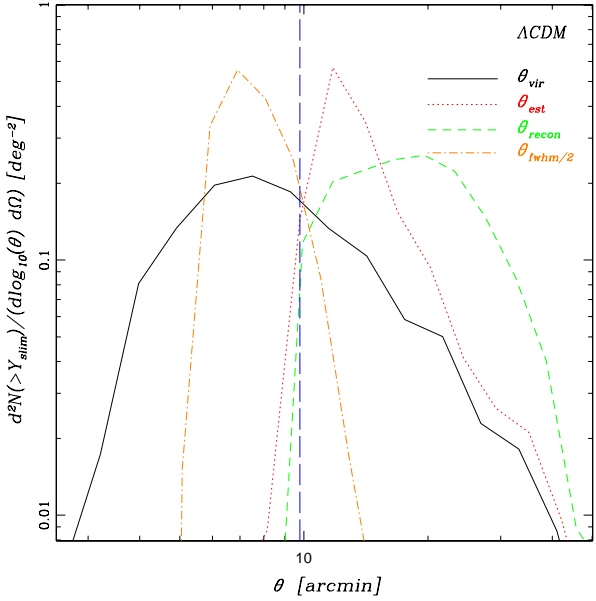}
\caption{The angular size distribution of clusters in the fiducial $\Lambda$CDM
survey with $Y\geq 2\times 10^{-3}$arcmin$^{2}$. The solid black curve represents
the distribution of the virial angular radius, the orange
dotted-dashed line the distribution of the radius at half the
amplitude of the cluster SZ signal after a convolution with a beam of
FWHM 9.7 arcmin. The distributions of the reconstructed
cluster radii and the expected radius reconstructions neglecting
cluster-cluster confusion
are given by
the green dashed and red dotted line respectively. The cluster is cut
off at the radius at which the change of the integrated flux with
radius falls below the contribution expected to be due to $3\times
\sigma_{noise}$. The blue
long-dashed vertical line represents the FWHM angular extent of the
beam estimate for the reconstructed map. \label{fig:rdist}}
\end{center}
\end{figure}

In Figure \ref{fig:rdist} we plot the distribution of the cluster angular
sizes for a plurality of cluster size definitions. First of all the
distribution of the virial radii obtained from the scaling relation
(\ref{eq:rv}) is shown. Furthermore, the radius distribution
of the beam-convolved
cluster extent at which the cluster SZ signal has dropped to half of
its peak amplitude is given. This can be compared with the FWHM of the
beam estimate of 9.7 arcmin for the HSMEM reconstructed SZ map. This
beam estimate of the
reconstructed thermal SZ map has been obtained by applying a $\chi ^2$-fit
of the convolved input map to the reconstructed thermal SZ map. If we
define that a cluster is convolved in the case that this FWHM radius
estimate exceeds the FWHM of the reconstruction beam, which ensures
that a cluster significantly contributes to several beams, a fraction
of approximately 10 per cent of the
recovered clusters of flux $Y \geq 2\times 10^{-3}$arcmin$^{2}$ is found to be
resolved.

Moreover, for comparison the expected and recovered cluster radius
distributions for the cluster extent at which the change of the
integrated flux with radius becomes equivalent to the contribution
expected from $3\times
\sigma_{noise}$ are shown in Figure \ref{fig:rdist}.
Both distributions show the same onset at low cluster radii
(approximately 8 to 9 arcmin).
Due to overestimation of the cluster radii by cluster-cluster
confusion and noise, the distribution obtained from the reconstructed SZ map
appears to be broader and less peaked. However, for large extended
clusters the reconstructed distribution again approaches the estimated
one and the algorithm provides thus a reasonable estimation of the cluster
radii. This occasional overestimation of projected small and mid-sized cluster radii
partially accounts for the
upscattering of the recovered flux distribution.

\subsection{The reconstructed SZ power spectrum}

\begin{figure}
\begin{center}
\includegraphics[scale=0.42]{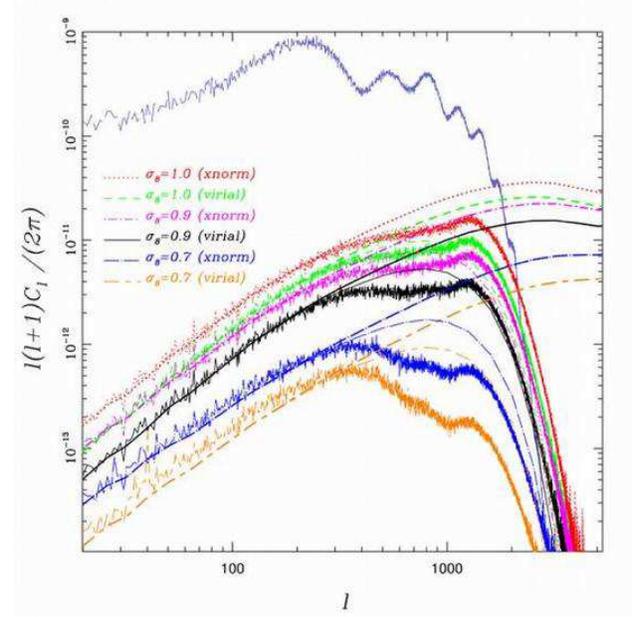}
\caption{Comparison of the reconstructed thermal SZ effect power
spectra in the case of the $\Lambda$CDM model ($\Omega_m=0.3$,
$\Lambda=0.7$, $h=0.7$; Table \ref{tab:param}) for different values of $\sigma_8$. The values
of $\sigma_8$, which refer to the particular input and reconstructed
power spectra, are indicated by the colour and line style. In addition
we plot the input (thick lines) and beam convolved (thin lines) SZ
power spectra. To obtain the beam convolved ones we assume a Gaussian
beam of FWHM of 9.7 arcmin. For comparison the reconstructed
primordial CMB power spectrum is
also shown. \label{fig:ps_lcdm_sz_recon}}
\end{center}
\end{figure}
\begin{figure}
\begin{center}
\includegraphics[scale=0.42]{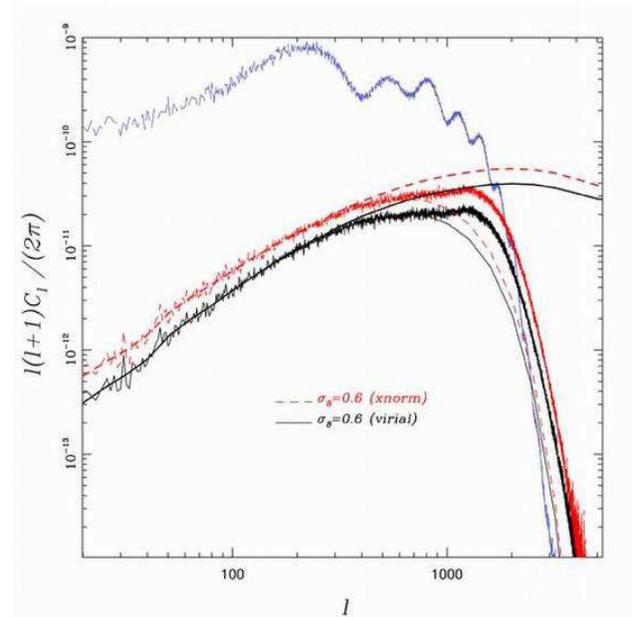}
\caption{The same as \ref{fig:ps_lcdm_sz_recon} for the $\tau$CDM
model ($\Omega_m=1$, $\Lambda=0$, $h=0.5$; Table \ref{tab:param}). \label{fig:ps_tcdm_sz_recon}}
\end{center}
\end{figure}

In addition to the cluster number count, the SZ power spectrum also strongly
depends on cosmology and can thus be used to constrain cosmological
parameters. Therefore, it is worthwhile investigating how well the SZ
power spectra of the different models can be reconstructed by the
HSMEM. In Figures \ref{fig:ps_lcdm_sz_recon} and \ref{fig:ps_tcdm_sz_recon} we
plot the HSMEM reconstructed SZ power spectra for each of the considered
models. For comparison also the input as well as the
beam-convolved SZ power spectra are shown. To obtain the beam-smoothed
power spectra, a beam of FWHM of 9.7 arcmin has been assumed, since it
provides the best $\chi^2$-fit to the reconstruction of our fiducial
$\Lambda$CDM model. 
Even though the HSMEM reconstructed power spectrum is intrinsically
biased (see Stolyarov et al.\,2002), in the case of the fiducial
$\Lambda$CDM model, the
reconstructed SZ power spectrum agrees well with the beam-convolved
input SZ power spectrum over a wide range of multipoles. Only on medium angular
scales ($500 < l < 1000$) the relative difference can be as large as
50 per cent. Nevertheless, at lower multipoles the reconstructed power
spectrum is in good agreement with the corresponding true one for most of the
models. Only in the case of the low-$\sigma_8$ $\Lambda$CDM model
(especially when using the virial M-T relation) the reconstructed
SZ power spectrum appears to be biased. In particular, the
reconstruction overpredicts the SZ power on large scales. This is due
to reconstruction and residual noise of other components present in the
reconstructed SZ map. Since the SZ power of the
low-$\sigma_8$ $\Lambda$CDM models is approximately an order of magnitude
smaller than for the high-$\sigma_8$
$\Lambda$CDM models, the obtained power spectra of the former models
are more affected by reconstruction and residual noise. Hence, one
obtains for these models an overestimated SZ power amplitude.

Furthermore, as the comparison of the beam-convolved
with the reconstructed power spectrum shows, models with larger
amplitudes than the fiducial $\Lambda$CDM model reproduce the true SZ power
spectrum better at higher multipoles and seem to be less
convolved. This effect arises
due to the relative increase of the SZ signal in comparison to the
instrumental noise and other
components with fixed amplitude, such as the Galactic dust, in the
high-resolution high-frequency
channels of Planck. The ones with lower
amplitudes show the opposite behaviour. However, up to multipoles of several hundreds the
reconstructed power spectra are not significantly affected by the
effective convolving beam of the reconstructed map. Thus given adequate prior
knowledge about the
spatial correlations of galaxy clusters, the outer cluster profile and
the M-T relation, our
simulations suggest that the reconstructed SZ power spectrum at low
multipoles can provide constraints on cosmological parameters in
addition to the cluster number count. 
As shown by Komatsu \& Seljak
(2002) the constraints on the cosmological parameters obtained by the
SZ power spectrum can be used to break degeneracies between parameter
constraints deduced from cluster number counts.

\section{Summary and Conclusions}
\label{sec:concl}

The Planck Surveyor satellite offers the opportunity to compile a large
sample of galaxy clusters over the full sky. This additional aim of
the Planck mission -- besides determining the primordial CMB power
spectrum at high angular resolution -- can be a very powerful
cosmological tool. We have performed full-sky multi-component simulations of
future Planck SZ observations with a high degree of realism. This is the first time that on the
basis of full-sky simulations and an application of a particular
detection algorithm the SZ cluster gain of the Planck
mission has been investigated. 

Based on cluster mass and redshift catalogues obtained
from N-body simulations and a mass function we derived a model sky, using further cluster
properties from scaling relations. Our modelling utilised two different
normalisations of the M-T scaling relation to explore the
dependence of our results on its choice. The one derived from
X-ray observations is more likely to include cluster physics, such
as preheating and cooling, and gives a higher gas temperature for a
given cluster mass. The theoretically
derived one has been normalised with
adiabatic hydrodynamical simulations and presents a lower bound. To model the extent of the
cluster electron gas we used the $\beta$-profile which is sufficient
given the resolution of Planck. On the basis of this modelling we
investigated the dependence of the SZ power spectra on $\sigma_8$. We
found that the power law index $\alpha(l)$ of the $C_{l}$ dependence on $\sigma_{8}$ possesses a negative slope in $l$-space,
which we argue can explain the wide range of estimates for
$\alpha$ obtained previously by various authors. Moreover, all
confusing extragalactic and Galactic components known to be
significant have been
incorporated in our full-sky simulations of the Planck channel
detections. By the current state of
art our modelling is sufficient to yield reliable
results.

Further, we have applied a SZ cluster detection algorithm existing of
two steps, the HSMEM to separate the SZ signal from other
contaminating components and a cluster finder.  In the present work
the simulated performance of Planck as a SZ cluster survey instrument
in combination with our detection algorithm has been tested for
several models including the ones favoured by current data. On the
basis of the fiducial $\Lambda$CDM cosmology (see Table
\ref{tab:param}) we estimated a conservative flux survey limit of $Y =
2\times  10^{-3}$arcmin$^{2}$. Given this limit a sufficient
completeness and purity is obtained for the majority of the
investigated models. Only in the case of the $\sigma_8=0.7$
$\Lambda$CDM model the completeness and purity have been found to be
significantly lower than 80\% for $Y_{surveylim} = 2\times
10^{-3}$arcmin$^{2}$. In the case that cluster contributions
to a single SZ peak detection can be disentangled, it has been
shown that the survey limit can be lowered to $1\times
10^{-3}$arcmin$^{2}$. Our investigations indicate that the cluster
selection of the detection algorithm is non-trivial and shows a
cosmological dependence mainly due to the strong cosmological
variation of the SZ signal magnitude. Ideally, the cluster detection
algorithm returns a cluster sample whose completeness and purity are
cosmology independent, since otherwise one has to rely on prior
information and mock simulations as performed in the present
work. However, in practice this ideal case can hardly be
reached. Nevertheless, prior constraints can be obtained from the
power spectrum of the SZ effect and thus allow a self-consistent
cosmological parameter estimation based on the Planck SZ
detection. Therefore, as suggested by our simulations besides serving
as a basis for follow-up requiring a major observational effort the
recovered Planck cluster catalogue itself
can already be used to put constraints on cosmological parameters. A
simple conservative investigation based on our SZ detection algorithm
in section \ref{sec:planckclnum} showed that a Planck SZ cluster
survey will be able to distinguish between the considered vanilla
models by at least a few $\sigma$. Given our algorithm we will
investigate how tightly cosmological parameters can be constrained by
a Planck SZ survey in a future paper.

Thus, as our results suggest Planck will not only be a powerful tool to
constrain cosmology via primordial CMB fluctuations but will also have
the ability to obtain
independently constraints from the secondary effect of SZ distortions. 

\section{Acknowledgements}

JG acknowledges support by an Isaac Newton Fellowship of the Isaac Newton Trust of the University of Cambridge. We especially thank Vlad Stolyarov for providing the HSMEM code and for several helpful discussions about its application to full-sky simulations.
Moreover, we acknowledge the HEALPix collaboration for providing the
pixelisation scheme used in this work and the Virgo Consortium for
making the Hubble Volume N-body simulations publically available.

\label{lastpage}

\end{document}